\documentclass{aa} 
\usepackage{amsmath} 
\usepackage{graphicx}
\usepackage{natbib}
\bibpunct{(}{)}{;}{a}{}{,}
\usepackage{txfonts}
\usepackage[T1]{fontenc}
\usepackage{ulem}
\usepackage{xcolor}
\newcommand{\cmt}{\,cm$^{-3}$}   
   
\newcommand{\kms}{\,km\,s$^{-1}$}

\newcommand{\myr}{\,$M_{\odot}\,{\rm yr}^{-1}$}

\newcommand{\ecs}{$\rm\,erg\,cm^{-2}\,s^{-1}$}

\newcommand{\ha}{H$\alpha$}
\newcommand{\hb}{H$\beta$}

\newcommand{\oIII}{[\ion{O}{iii}]}

\newcommand{\ro}{\,$R_{\odot}$}

\newcommand{\lo}{\,$L_{\odot}$}
\begin{document}

%
%
\title{Wind mass transfer in S-type symbiotic binaries\thanks{Tables 1 and 2 are only available 
in electronic form at the CDS via anonymous ftp to 
{\tt cdsarc.u-strasbg.fr} ({\tt 130.79.128.5}) or via 
{\tt http://cdsweb.u-strasbg.fr/cgi-bin/qcat?J/A+A/}}}
\subtitle{III. Confirmation of a wind focusing in EG Andromedae\\ 
          from the nebular \oIII\,$\lambda$5007 line}
\titlerunning{Wind focusing in EG Andromedae}
\authorrunning{N.~Shagatova, A.~Skopal, S.~Yu.~Shugarov et al. }
\author{N.~Shagatova    \inst{\ref{inst1}}
  \and A.~Skopal        \inst{\ref{inst1}}
  \and S.~Yu.~Shugarov  \inst{\ref{inst1},\ref{inst2}}
  \and R.~Kom\v z\' ik  \inst{\ref{inst1}} 
  \and E.~Kundra        \inst{\ref{inst1}}
  \and F.~Teyssier      \inst{\ref{inst3}}
}
\institute{Astronomical Institute, Slovak Academy of Sciences,
           059~60 Tatransk\'{a} Lomnica, Slovakia
           \email{nshagatova@ta3.sk}\label{inst1} 
      \and P.K. Sternberg Astronomical Institute, M.V. Lomonosov 
           Moscow State University, Russia\label{inst2}
      \and Observatoire Rouen Sud (FR) - ARAS group, 76100 Rouen, 
           France\label{inst3}
}

\date{Received / Accepted }

\abstract
{
The structure of the wind from the cool giants in symbiotic binaries 
carries important information for understanding the wind mass 
transfer to their white dwarf companions, its fuelling, and 
thus the path towards different phases of symbiotic-star evolution. 
}
{
In this paper, we indicate a non-spherical distribution of the 
neutral wind zone around the red giant (RG) in the symbiotic 
binary star, \object{EG And}. We concentrate in particular on 
the wind focusing towards the orbital 
plane and its asymmetry alongside the orbital motion of the RG. 
}
{
We achieved this aim by analysing the periodic orbital variations 
of fluxes and radial velocities of individual components of the 
\ha\ and [\ion{O}{iii}]\,$\lambda$5007 lines observed on our 
high-cadence medium ($R\sim 11\,000$) and high-resolution 
($R\sim 38\,000$) spectra. 
}
{
The asymmetric shaping of the neutral wind zone at 
the near-orbital-plane region 
is indicated by: 
(i) the asymmetric course of the \ha \ core emission fluxes along 
    the orbit; 
(ii) the presence of their secondary maximum around the orbital 
     phase $\varphi = 0.1, $ which is possibly caused by the refraction 
     effect; and 
(iii) the properties of the \ha\ broad wing emission originating 
      by Raman scattering on H$^0$ atoms.
The wind is substantially compressed from polar directions 
to the orbital plane as constrained by the location of the 
[\ion{O}{iii}]\,$\lambda$5007 line emission zones in the 
vicinity of the RG at/around its poles. 
The corresponding mass-loss rate from the polar regions of 
$\lesssim$10$^{-8}$\myr\ is a factor of $\gtrsim$10 lower than 
the average rate of $\approx$10$^{-7}$\myr\ derived from nebular 
emission of the ionised wind from the RG. Furthermore, it is two orders 
of magnitude lower than that measured in the near-orbital-plane 
region from Rayleigh scattering. 
}
{
The startling properties of the nebular [\ion{O}{iii}]\,$\lambda$5007 
line in \object{EG And} provides an independent indication of the wind 
focusing towards the orbital plane -- the key to understanding 
the efficient wind mass transfer in symbiotic binary stars. 
}
\maketitle
\section{Introduction}
\label{s:int}
Symbiotic stars (SySts) are the widest interacting binary systems 
comprising a cool giant as the donor star and a hot compact star, 
mostly a white dwarf (WD), accreting from the giant's wind 
\citep[][]{ms99}. 
On the basis of their infrared properties, we distinguish 
between S-type (Stellar) and D-type (Dusty) SySts 
\citep[][]{1975MNRAS.171..171W}. 
S-type systems comprise a normal red giant (RG) with typical 
orbital periods of a few years \citep[e.g.][]{be+00}, while 
D-type systems contain a Mira-type variable and their orbital 
periods are considered to be of a few times 10--100 years 
\citep[e.g.][]{mk06}.
The accretion process heats up the WD to $\ga 10^5$\,K and 
increases its luminosity to $10^1 - 10^4$\lo\ \citep[][]{mu+91,sk05}. 
The accreting material converts its gravitational potential energy 
into the radiation by a disk \citep[e.g.][]{1981ARA&A..19..137P}, 
but it can also serve as a supply for stable nuclear burning on the WD surface
\citep[e.g.][]{1980A&A....82..349P}. 
In the former, case we talk about accretion-powered  
systems, whose luminosities are as low as a few times 10\lo, while 
the latter are nuclear-powered systems  
generating luminosities of a few times 10$^3$\lo. 
Consequently, the accreting WD ionises a fraction of the wind from 
the RG dividing the binary environment into the ionised zone 
around the WD and the neutral zone around the giant, both limited 
by the common H$^{0}$/H$^{+}$ boundary \citep[e.g.][]{se+84,nv87}. 
This configuration represents the so-called quiescent phase. 
Sometimes, SySts undergo outbursts characterised by a few magnitude 
brightening(s) observed on a timescale of a few months to 
years/decades and signatures of the enhanced mass outflow 
\citep[e.g.][]{mu19}. 

The principal problem in the field of SySt research 
inheres in the large energetic output from burning 
WDs and the too-low accretion efficiency via the wind mass 
transfer in the canonical Bondi-Hoyle type of accretion 
\citep[][]{kg83}. 
The solution to this problem requires a more effective mass transfer 
resulting from a non-spherically symmetric wind of the donor star. 
For a D-type system, the binary star \object{Mira} ($o$ Ceti), observations 
revealed a highly aspherical circumbinary environment with 
a mass outflow from Mira towards the WD \citep[see][]{ma+01,ka+04}. 
Accordingly, \cite{mp07} suggested an efficient wind mass 
transfer mode for Mira-type interacting binaries, whose slow 
and dense wind from the evolved star on the asymptotic giant branch (AGB) is filling the Roche 
lobe instead of the star itself: the so-called wind Roche-lobe 
overflow \citep[see also][]{2013A&A...552A..26A}. 
For S-type system \object{SY Mus}, \cite{du+99} evidenced 
an asymmetric density distribution around the RG by measuring 
the hydrogen column densities, which was later confirmed by modelling its 
UV light curves by \cite{ss17}. 
Furthermore, \cite{sh+16} indicated a compression of the RG's 
wind towards the orbital plane in the symbiotic binary \object{EG And} 
and \object{SY Mus} by modelling the H$^0$ column densities 
measured from Rayleigh scattering. 
A possible solution to the wind mass-transfer problem in the S-type 
SySts was suggested by \cite{2015A&A...573A...8S}, who applied 
the wind compression disk model of \cite{bc93} to slowly rotating 
RGs in S-type systems 
\citep[][]{2007MNRAS.380.1053Z,2008MNRAS.390..377Z}, which 
leads to the enhancement of the wind mass loss in the orbital plane. 
The primary aim of this paper is to demostrate a 
focusing of the wind from the RG in the SySt \object{EG And} 
independently from the previous approach of \cite{sh+16}. 

The accretion-powered S-type symbiotic 
binary \object{EG And} has no history of outbursts. The binary comprises 
an M3.5\,{\small III} RG and a WD accreting from 
the giant's wind on a 483-d orbit \citep[e.g.][]{fe+00} with 
a high inclination of $\approx 80\degr$ \citep[][]{vo+92}. 
An accretion process heats the WD to a temperature of 
$\approx 7.5 - 9.5\times 10^4$\,K \citep{kg16,si+17}, 
producing the luminosity of a few times 10\lo\ for a distance of 
$0.3 - 0.6$\,kpc
\citep[e.g.][]{mu+91,sk05}. 
Therefore, \object{EG And} is a low-excitation SySt with a faint nebular 
continuum and a few emission lines (mostly \ion{H}{i} and some 
forbidden lines) superposed on the RG continuum in 
the optical. 
The quiescent phase configuration of \object{EG And} and its high 
orbital inclination induce pronounced spectral variations along 
the orbit 
\citep[e.g.][]{sm80,1984ApJ...281L..75S,ol+85,sk+91,mu93,
1996Ap&SS.235..305P,kg16,ko+18}, which is best seen in the \ha\ line 
profile.

The true nature of the pronounced orbitally related variation 
in the \ha\ profile of quiet SySts is not well understood 
to date. 
For \object{EG And}, some authors \citep[e.g.][]{to95,it04,ca14} attempted 
to explain this variability within the theory of colliding winds 
in interacting binaries suggested by \cite{gw87} and \cite{lu+90}. 
However, the fact that a fraction of the giant's wind has to be 
transferred to and accreted onto the WD 
excludes the presence of colliding winds, at least 
at/around the orbital plane \citep[see also][]{wf00,kg16}. 
On the other hand, the presence of the faint X-ray spectrum of \object{EG And}, 
which is of the $\beta$ type \citep[][]{mu+97},  
is understood to be a result of a shock-heated plasma from 
colliding winds \citep[see][]{mu+95}. These facts 
constrain a biconical structure of the wind from the accreting WD, 
which can collide with the RG wind well above the orbital plane. 
On the basis of 20 years of spectroscopic observations, 
\cite{kg16} confirmed the variations of \ha\ and \hb\ equivalent 
widths along the orbit that are consistent with the apparent 
changes of the \ion{H}{ii} region arising from the ionisation of 
the RG wind by the hot component. 
%
A similar orbital modulation of the \ha\ line profile was also observed for quiescent, nuclear-powered eclipsing SySt \object{SY Mus} 
and \object{RW Hya} \citep[see][]{sc+94,sc+96}. The authors suggested    
that the bulk of the \ha\ emission lies in a small high-density
region above the RG surface facing the hot component. For \object{RW Hya},
they observed that the broad wings are subject to the eclipse. 

To obtain information on the RG wind distribution 
in the symbiotic binary \object{EG And}, we investigated the orbitally 
related variation in the \ha\ and the nebular 
[\ion{O}{iii}]\,$\lambda$5007 line profiles, because these lines 
are created under very different physical conditions. 
In Sect.~\ref{s:obs}, we describe our observations and their 
treatment, while in Sect.~\ref{anres} we present our analysis and 
results. The discussion and summary can be found in 
Sects.~\ref{s:dis}. and \ref{s:sum}. 

\begin{table}[t!]
   \caption{
Log of spectroscopic observations. The full table is available 
at the CDS. 
HJD stands for heliocentric Julian date. 
           }
\label{tab:log}
\centering
\begin{tabular}{cccc}
\hline
\hline
\noalign{\smallskip}
  HJD  &      Date      & line$^{a)}$& Obs.$^{b)}$  \\
       & (yyyy/mm/dd.ddd) &           &             \\
\noalign{\smallskip}
\hline
\noalign{\smallskip}
 2457256.436  & 2015/08/21.936  & \ha & ARAS   \\
 2457258.510  & 2015/08/24.010  & \ha & G1     \\
 2457263.477  & 2015/08/28.977  & \ha & ARAS   \\
   ...        &    ...          &     & ..     \\
 2459163.305  & 2020/11/09.805  & [\ion{O}{iii}], \ha & SP\\
\noalign{\smallskip}
\hline
\end{tabular}
{\bf Notes.} 
$^{a)}$\,analysed lines: [\ion{O}{iii}]$\lambda5007$, \ha, 
$^{b)}$\,Observatory
\end{table}

\section{Observations}
\label{s:obs}

To follow the evolution of the selected emission line profiles, 
we used 120 high- and medium-resolution optical spectra 
(Table~\ref{tab:log}), from which 109 and 33 were used 
to analyse the \ha\ and the [\ion{O}{iii}]\,$\lambda$5007 
composite line profile, respectively. 
The spectra were obtained during years 2015--2020 by telescopes 
at the Star\'a Lesn\'a (G1 in Table~\ref{tab:log}) and 
Skalnat\'e Pleso (SP) observatories, complemented with those 
available at the Astronomical Ring for Access to Spectroscopy 
database (ARAS;\footnote{\tt 
www.astrosurf.com/aras/Aras\_DataBase/Symbiotics.htm}  \cite{te19}). 

The G1 observatory is equipped with a 0.6\,m Cassegrain telescope 
(f/12.5) with an eShel fibre-fed spectrograph (R$\sim$11\,000), and 
the SP observatory has a 1.3\,m Nasmyth-Cassegrain telescope 
(f/8.36) with a fibre-fed \'{e}chelle spectrograph (R$\sim$38\,000) 
similar to the MUSICOS design \citep[][]{bb92,pr+15}. 
The G1 and SP spectra were reduced with the Image Reduction and 
Analysis Facility (IRAF;\footnote{\tt http://iraf.noao.edu}  
\cite{to86}) using specific scripts and programs 
\citep[see][]{pr+15}. The calibration is performed using 
the unit containing the ThAr hollow cathode lamp, 
tungsten lamp, and blue LED. 

The ARAS medium-resolution spectra were obtained at the private 
station in Rouen with a 0.36\,m Schmidt-Cassegrain telescope (F/D = 5) 
equipped with an Echelle spectrograph (R$\sim$11\,000) and 
ATIK 460Ex CCD camera $ (2\times 2$ binning). 
 One more ARAS spectrum was obtained at the private station, 
 Mill Ridge, with a 0.31\,m Cassegrain telescope equipped with 
 a classical slit spectrograph. The detector was a QSI583 (KAF8300
chip) binned $1\times 2$ (R$\sim$9000).
We also used two low-resolution ARAS 
spectra (Fig.~\ref{gscale}). One taken at the orbital phase 
$\varphi = 0$ (obtained 
by J. Guarro using a Control Remote Telescope SC16, spectrograph 
B60050-VI, and an ATIK 460EX CCD camera) and the second one at the 
opposite phase, $\varphi = 0.5$ (obtained by Martineau-Buchet 
using a Celestron C11 telescope, LISA spectrograph, and a QSI583 
wsg CCD camera). 
The ARAS spectra were processed and merged using Integrated 
Spectrographic Innovative Software (ISIS).\footnote{\tt 
www.astrosurf.com/buil/isis-software.html}
The wavelength calibration is performed using the spectrum of 
a Thorium lamp.  

\begin{figure}
\centering
\begin{center}
\resizebox{6cm}{!}{\includegraphics[angle=-90]{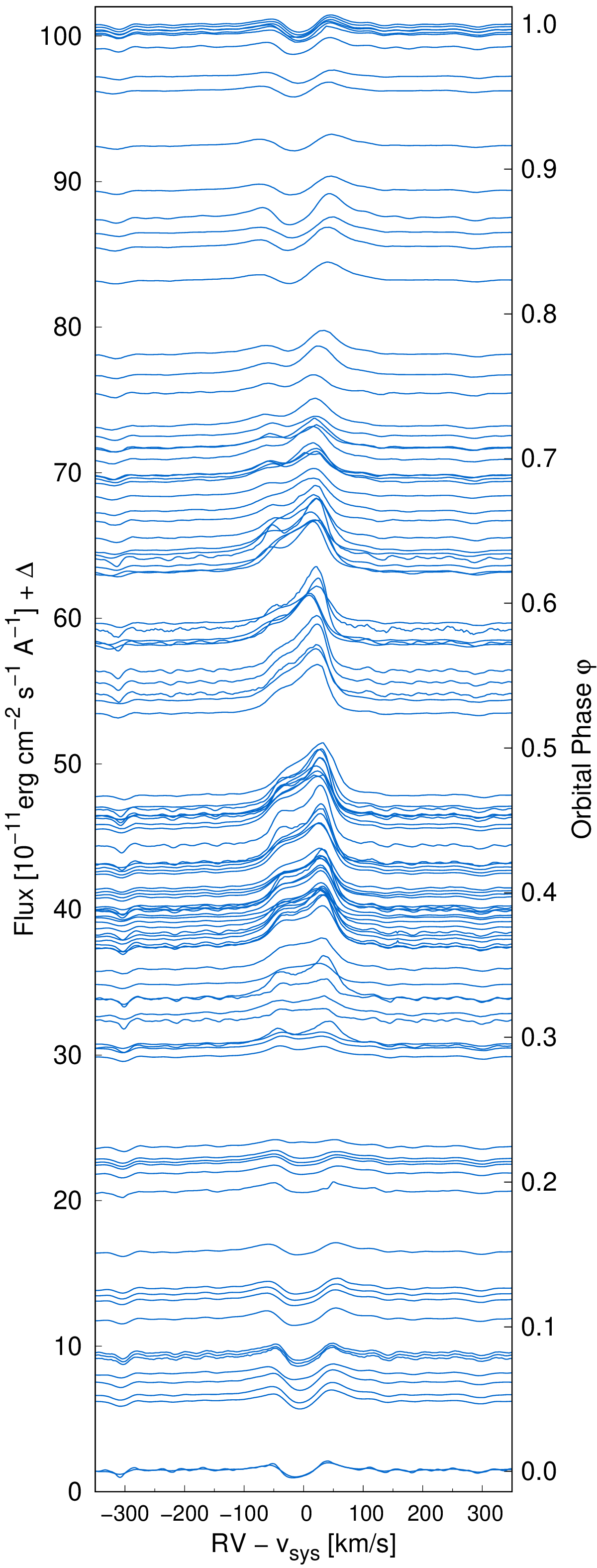}}
\end{center}
\caption{
Set of 109 spectra of \object{EG And} around the \ha\ line, shifted by 
$\Delta = 100\times \varphi$ in fluxes to demonstrate the orbital 
modulation. 
        }
\label{fig:allha}
\end{figure}

All spectra were dereddened with $E_{\rm B-V} = 0.05$\,mag 
\citep[][]{mu+91} using the extinction curve of \cite{ca+89}. 
We determined the orbital phase of the binary according to 
the ephemeris of the inferior conjunction of the RG 
($\varphi = 0$) as follows \citep[see][]{fe+00,kg16}: 
\begin{equation}
 JD_{\rm sp. conj.} = 
    2\,450\,683.2(\pm 2.4) + 482.6(\pm 0.5)\times E .
\end{equation}
Moreover, we used the systemic velocity of \object{EG And}, 
$v_{\rm sys} = -94.88$\,kms$^{-1}$, as derived by \cite{kg16} 
using the SAO radial velocity data, which agree with those 
determined by other authors \citep{ol+85,mu+88,mu93,fe+00}. 
Figure~\ref{fig:allha} shows a \ha\ profile in all our spectra 
as a function of $\varphi$ in the radial velocity space, 
$RV - RV_{\rm sys}$. 
%
%
%
\section{Analysis and results}
\label{anres}
\subsection{Determination of the true continuum}
\label{ss:cont}

\begin{figure}
\centering
\begin{center}
\resizebox{\hsize}{!}{\includegraphics[angle=-90]{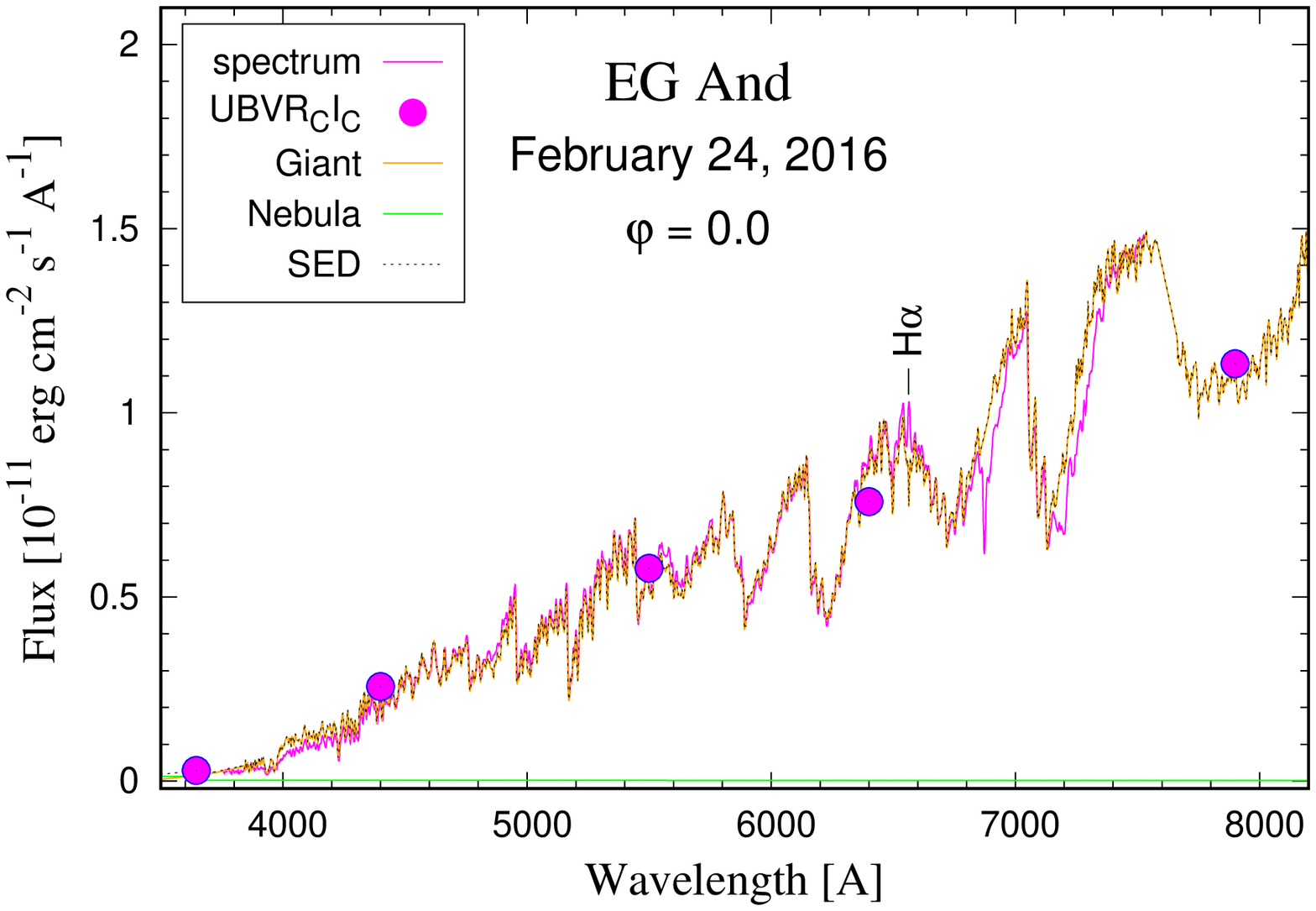}}
\resizebox{\hsize}{!}{\includegraphics[angle=-90]{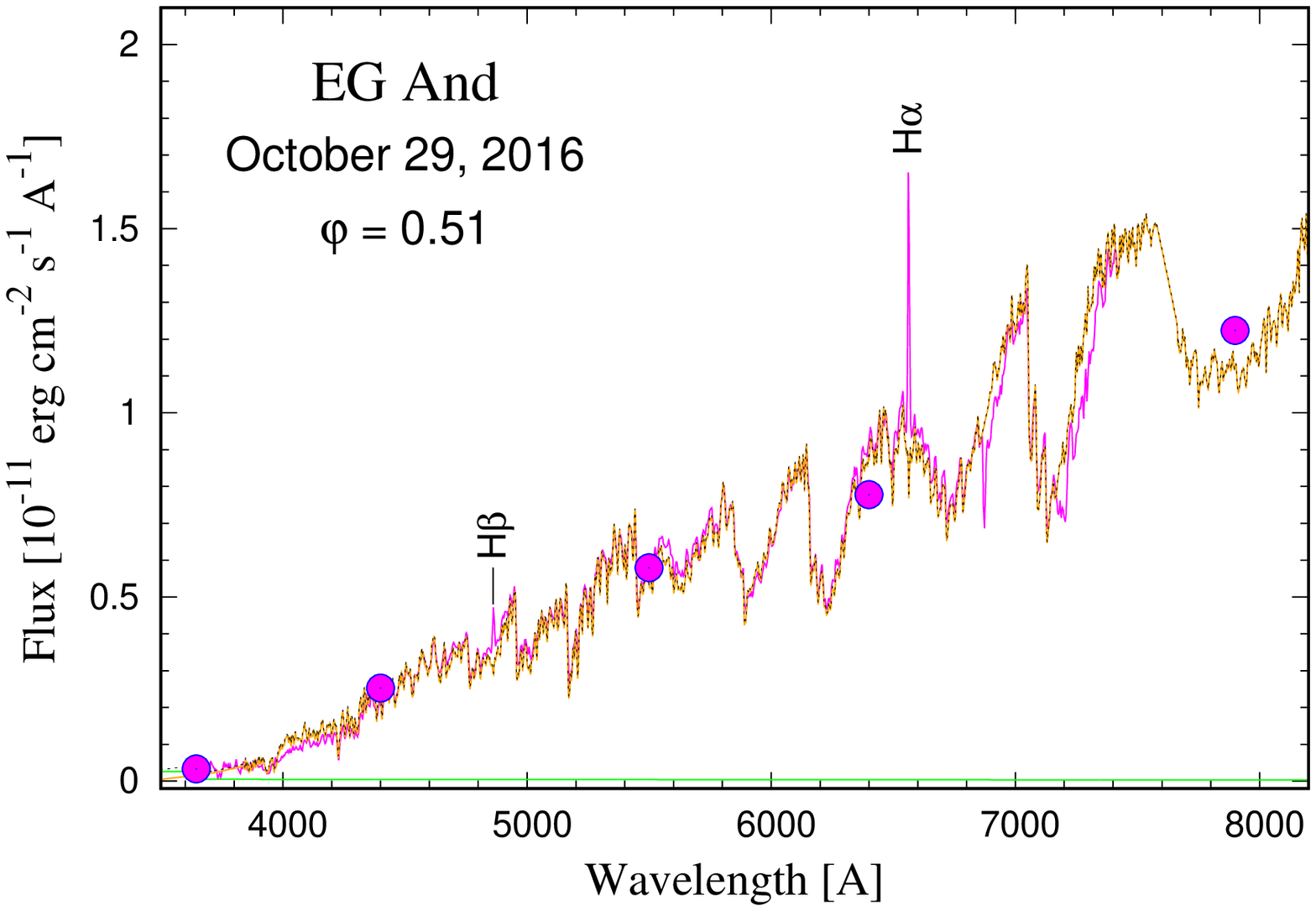}}
\end{center}
\caption{
Low-resolution spectra of \object{EG And} obtained at the orbital 
phase $\varphi = 0$ (top) and $\varphi = 0.51$ (bottom), 
demonstrating significant variations in fluxes of hydrogen 
lines with the orbit. 
The spectrum is dominated by a M3.4\,{\small III} RG, while 
the nebular continuum is very small, and the contribution from 
the hot WD cannot be determined in the optical 
\citep[for details, see Sect.~3.2.1. of][]{sk+17}. 
        }
\label{gscale}
\end{figure}

To convert the spectra from relative to absolute fluxes, we used 
the $UBVR_{\rm C}$ photometry of \object{EG And} as published by \cite{sk+12} 
and \cite{2019CoSka..49...19S}. We converted the photometric 
magnitudes to fluxes according to the calibration of \cite{hk82}. 
Then, we determined the level of the true continuum by 
scaling the spectrum at given orbital phase to the average 
photometric fluxes measured at the same $\varphi$. 
In addition, by modelling the low-resolution spectra 
 \citep[see Sect.~3.2.1. of][]{sk+17} we 
obtained the spectral type of the RG as M3.4\,{\small III} 
(see Fig.~\ref{gscale}) using 
the models of \cite{fl+94}. 
The corresponding synthetic spectrum then determines 
the profile of the true continuum of our medium- and 
high-resolution spectra (see Figs.~\ref{fitsHa} 
and \ref{fig:oiii}). 
%
%
%
%
%
\subsection{Disentangling the \ha\ line profile}
\label{ss:ha}
\begin{figure}
\centering
\begin{center}
\resizebox{\hsize}{!}{\includegraphics[angle=0]{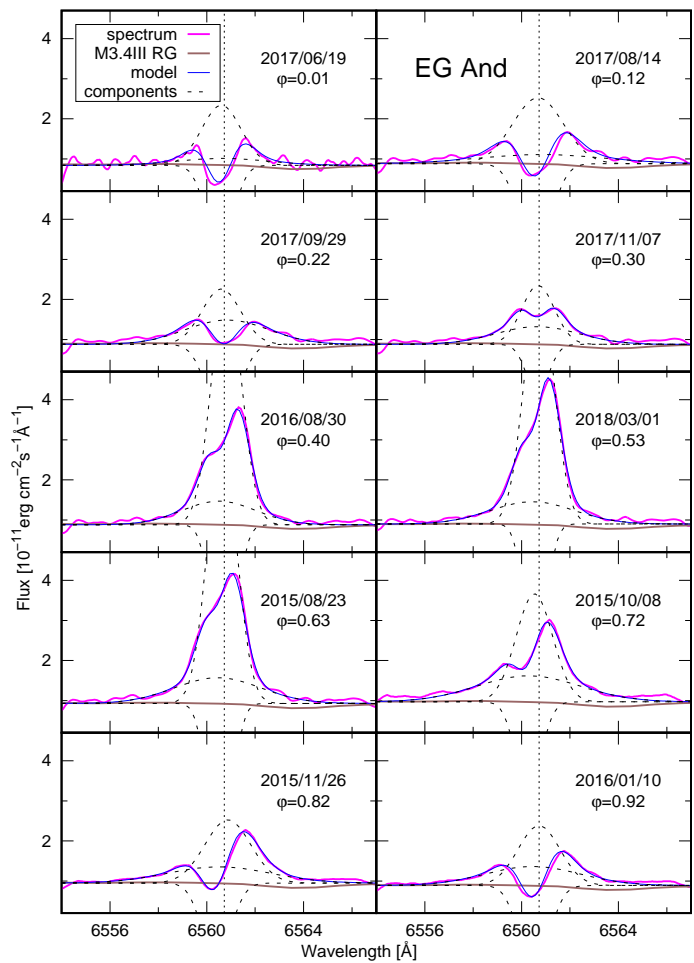}}
\end{center}
\caption{
Example of observed \ha\ line profile, its fit by 
Eq.~(\ref{eq:ha}), and its individual components (Sect.~\ref{ss:ha}) 
at different orbital phases. Vertical dotted line corresponds 
to the wavelength of the systemic velocity. 
        }
\label{fitsHa}
\end{figure}

Assuming that the \ha\ line consists of three basic components, 
the core emission, 
$G_{\rm CE}(\lambda)$, the broad wings emission, 
$G_{\rm WE}(\lambda)$, and the absorption, 
$G_{\rm A}(\lambda)$, 
its profile can be expressed as 
\begin{equation}
  P_{\rm H\alpha}(\lambda) = G_{\rm CE}(\lambda)
                           + G_{\rm WE}(\lambda)
                           + G_{\rm A}(\lambda).
\label{eq:ha}
\end{equation}
Furthermore, we approximated each component with the following Gaussian function: 
\begin{equation}
  G(\lambda) = I\exp\left[-\ln(2)\left(\frac{\lambda-\lambda_0}{h}
              \right)^2 \right],
\label{eq:gauss}
\end{equation}
where $I$ is the scaling, $\lambda_0$ is the central wavelength,
and $h$ is the half width at half maximum of the given component. 
To determine these parameters for each component, we fitted 
the \ha\ line profile with the function (\ref{eq:ha}) using 
the code Fityk.\footnote{\tt https://fityk.nieto.pl/} 
Figure~\ref{fitsHa} shows a preview of selected fits 
at different orbital phases, which demonstrates the variability 
of the \ha\ line profile along the orbit. 
The wavelength $\lambda_0$ and parameters $I$, $h$ determine the 
RV and the flux $(F = I h \sqrt(\pi/\ln(2))$) of the given 
component. Values of these parameters for each component 
of the \ha\ profile of all our spectra are introduced in 
Table~\ref{tab:par} and depicted in Fig.~\ref{fig:flx_rvs}. 
%
%
%
\begin{table*}[t!]
   \caption{
Fluxes and RVs of individual components of \ha\ line profile.
The full table 
is available at the CDS. 
           }
\label{tab:par}
\centering
\begin{tabular}{cccccccc}
\hline
\hline
\noalign{\smallskip}
Date                & 
Phase               &
$F_{\rm CE}$        & 
$F_{\rm WE}$        & 
$F_{\rm A}$         & 
$RV_{\rm CE}$       & 
$RV_{\rm WE}$       & 
$RV_{\rm A}$        \\ 
(yyyy/mm/dd.ddd)                     &
                                     &
\multicolumn{3}{c}{[$10^{-11}$\ecs]} &
\multicolumn{3}{c}{[\kms]}           \\
\noalign{\smallskip}
\hline
\noalign{\smallskip}
2015/08/21.936  & 0.621 & 10.652  & 3.617 & 4.769 &  -98.670 &  -98.670 & -102.781 \\
2015/08/24.010  & 0.626 &  9.834  & 2.858 & 5.129 & -100.040 & -108.263 & -104.151 \\
2015/08/28.977  & 0.636 &  9.978  & 4.492 & 4.660 & -100.040 & -106.435 & -106.435 \\
     ...        &  ...  &   ...   &  ...  & ...   &    ...   &    ...   &    ...   \\
2020/11/09.805  & 0.573 & 19.413  & 3.807 & 15.170 &  -98.213 & -106.892 & -100.954 \\
\noalign{\smallskip}
\hline
\end{tabular}
\end{table*}

We evaluated the goodness of fits using the reduced  
$\chi_{\rm red}^2$ quantity. 
For all fits, $\chi_{\rm red}^2 > 1$, 
which corresponds to a standard deviation of fluxes of $2\%$. 
Our fitting procedure provides the errors of heights $I$ that are often 
larger than 10\%.  
This is because we did not allow the height of the CE component 
to vary freely, otherwise the fitting procedure would have converged 
to its unrealistically high values due to the degeneracy between 
the CE and A component. Therefore, we caution that our flux values 
can be considered only as lower limits. 
To verify the overall trend in the CE and A component fluxes, 
we plot their ratio in  
Fig.~\ref{fig:fluxCEoverA}, which 
shows the maximum at $\varphi \sim 0.4$ and 
the secondary maximum at $\varphi\sim 0.1$. 
Finally, the uncertainties of fluxes of the disentangled components 
can be even larger due to calibration errors and the simplicity of our 
approach. Therefore, the resulting fits do not `perfectly' match 
the observed \ha\ line profile (see Fig. \ref{fitsHa}). 

With regard to the $RV$ values, our fitting procedure and the 
resolution of our spectra allowed us to estimate their 
uncertainties to typically $\pm (1-5)$ and $\pm (1-10)$\kms\ 
for the CE/A and the broad WE component of the \ha\ profile. 
For the [\ion{O}{iii}]\,$\lambda$5007 line, the uncertainties 
in RVs are estimated as $\pm (0.2-0.7)$ and $\pm (0.5-1)$\kms\ 
for the medium- and high-resolution spectra, respectively. 

\subsection{Fluxes of the \ha\ line components}
\label{ss:fha}

\begin{figure*}[]
\centering
\begin{center}

\resizebox{\hsize}{!}{\includegraphics[angle=0]{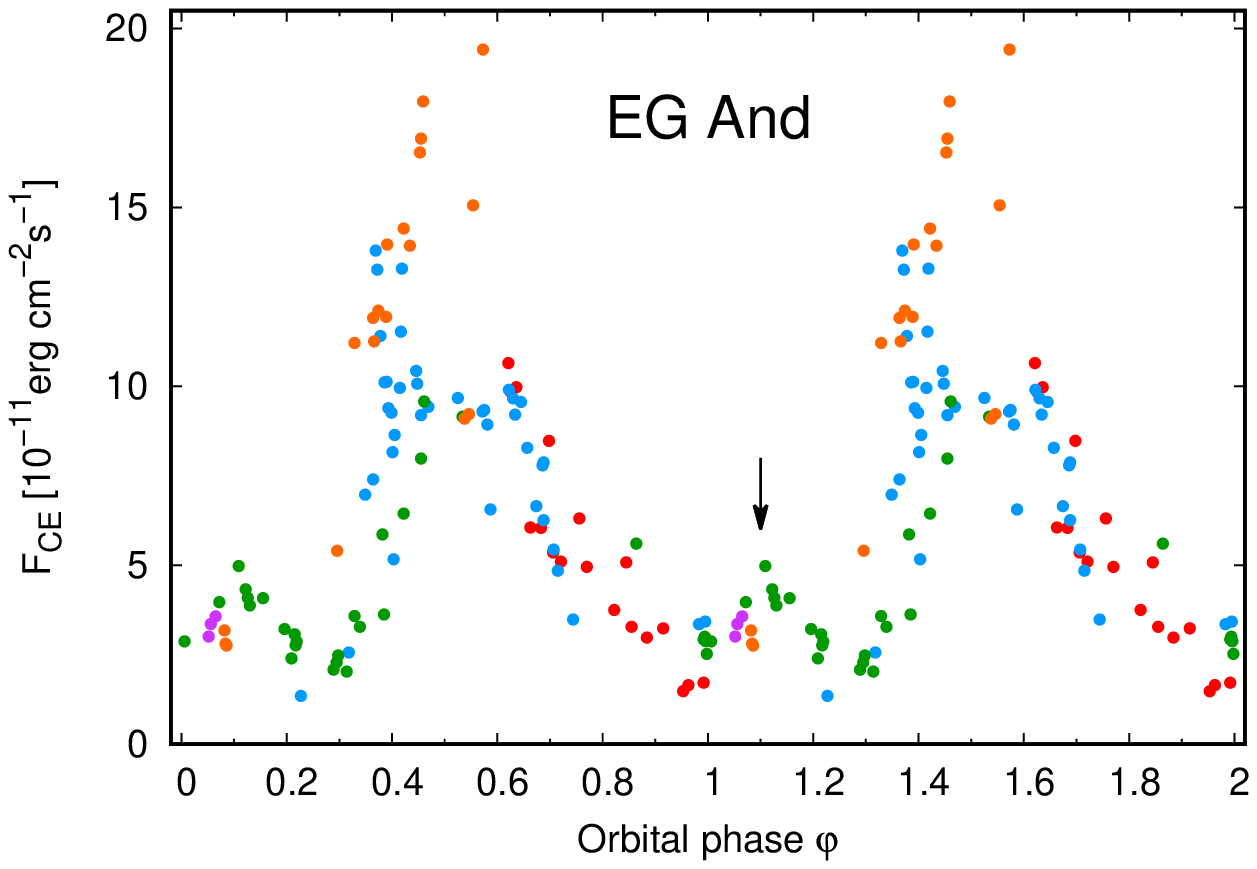}
                      \includegraphics[angle=0]{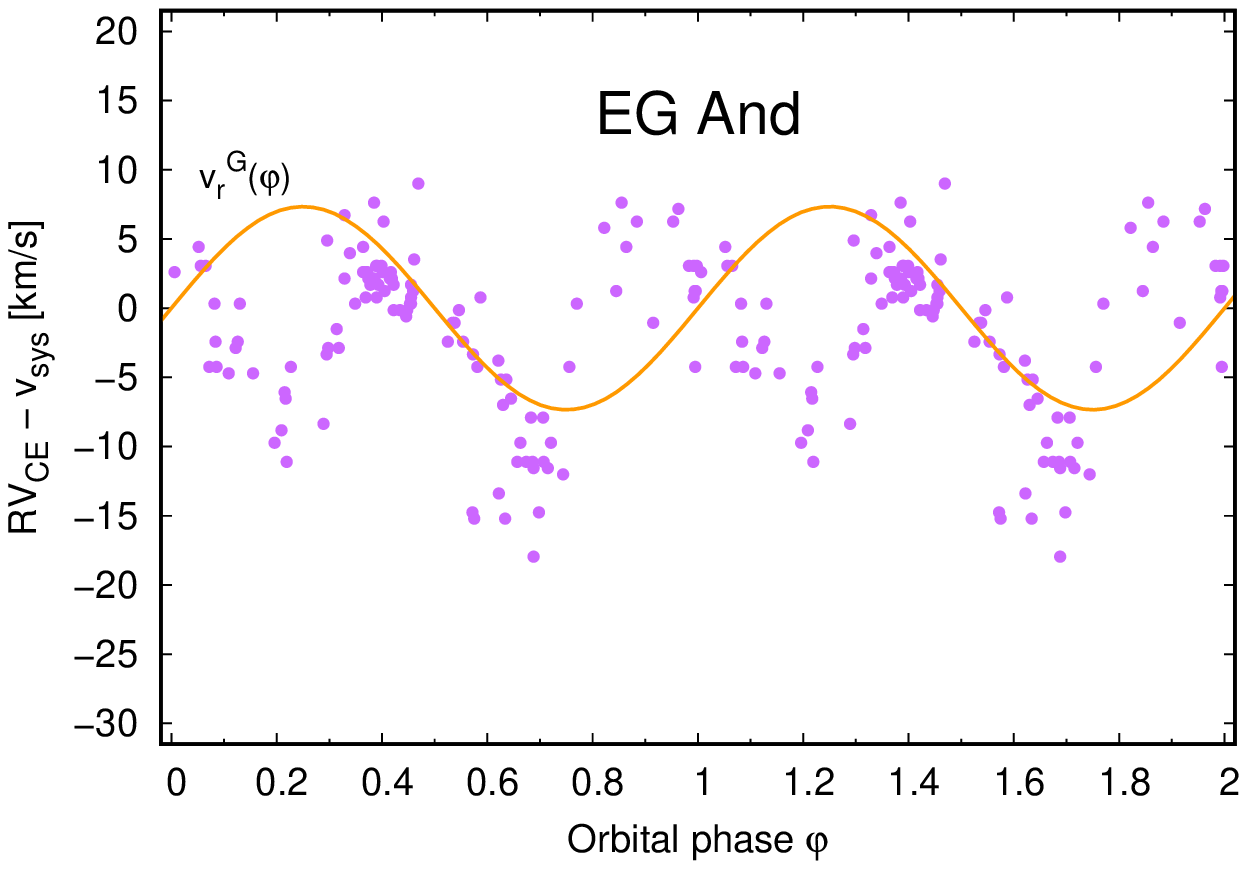}}
\resizebox{\hsize}{!}{\includegraphics[angle=0]{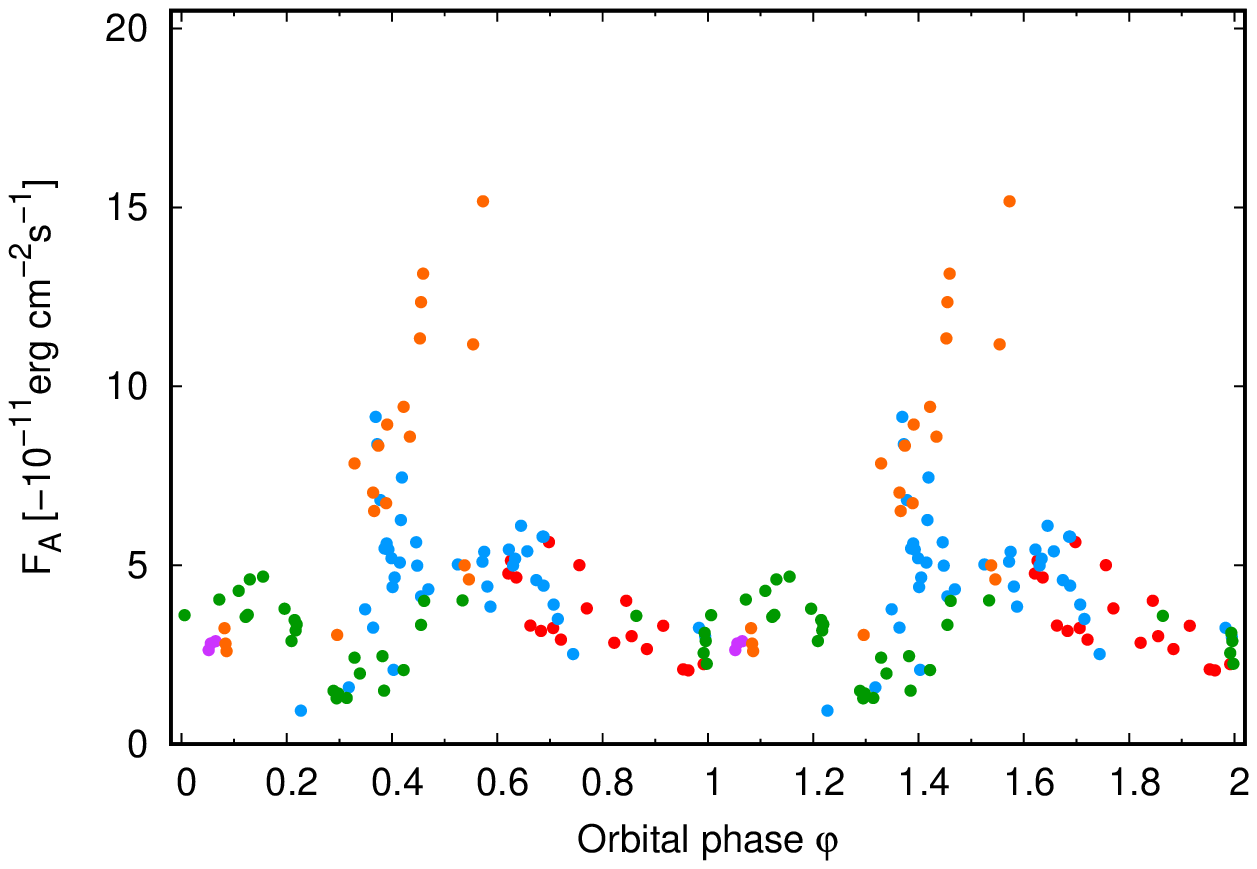}
                      \includegraphics[angle=0]{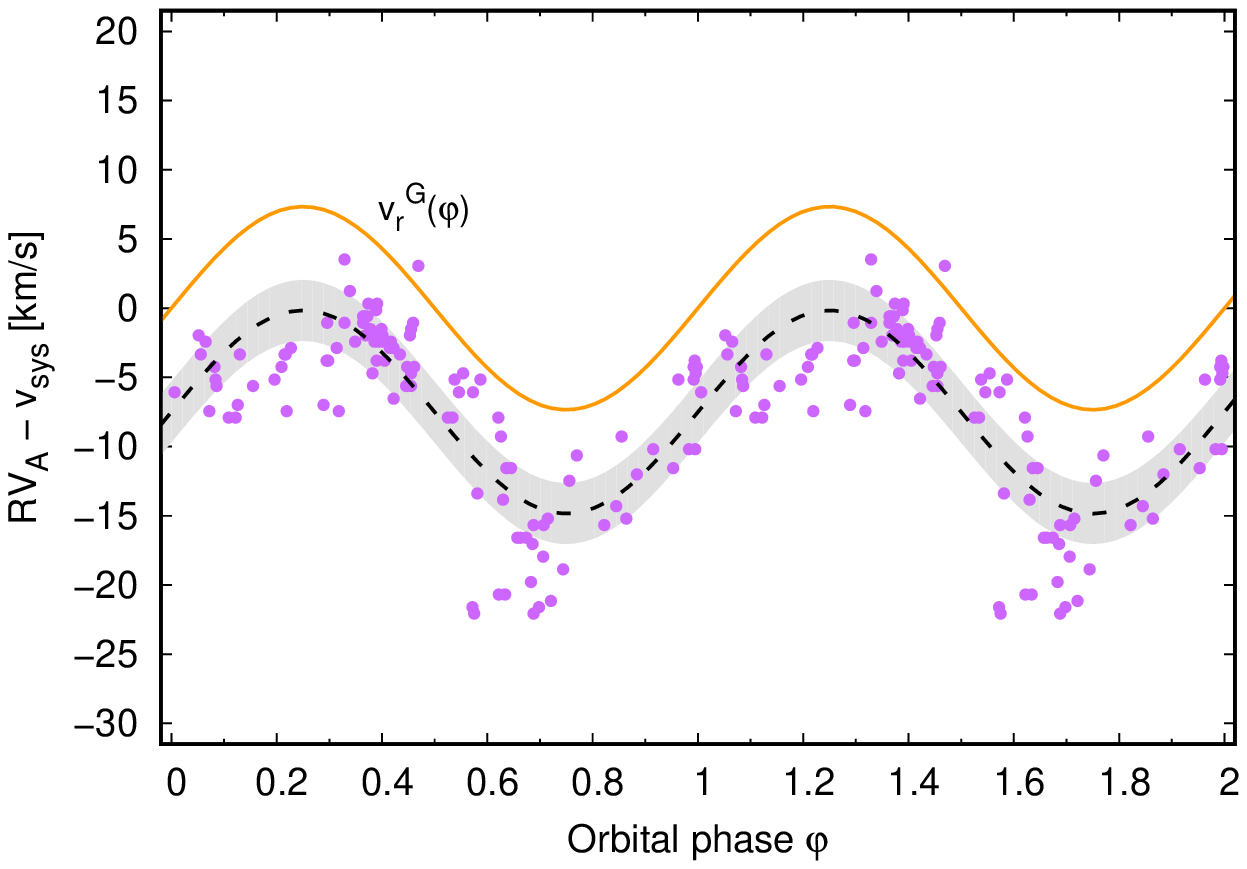}}
\resizebox{\hsize}{!}{\includegraphics[angle=0]{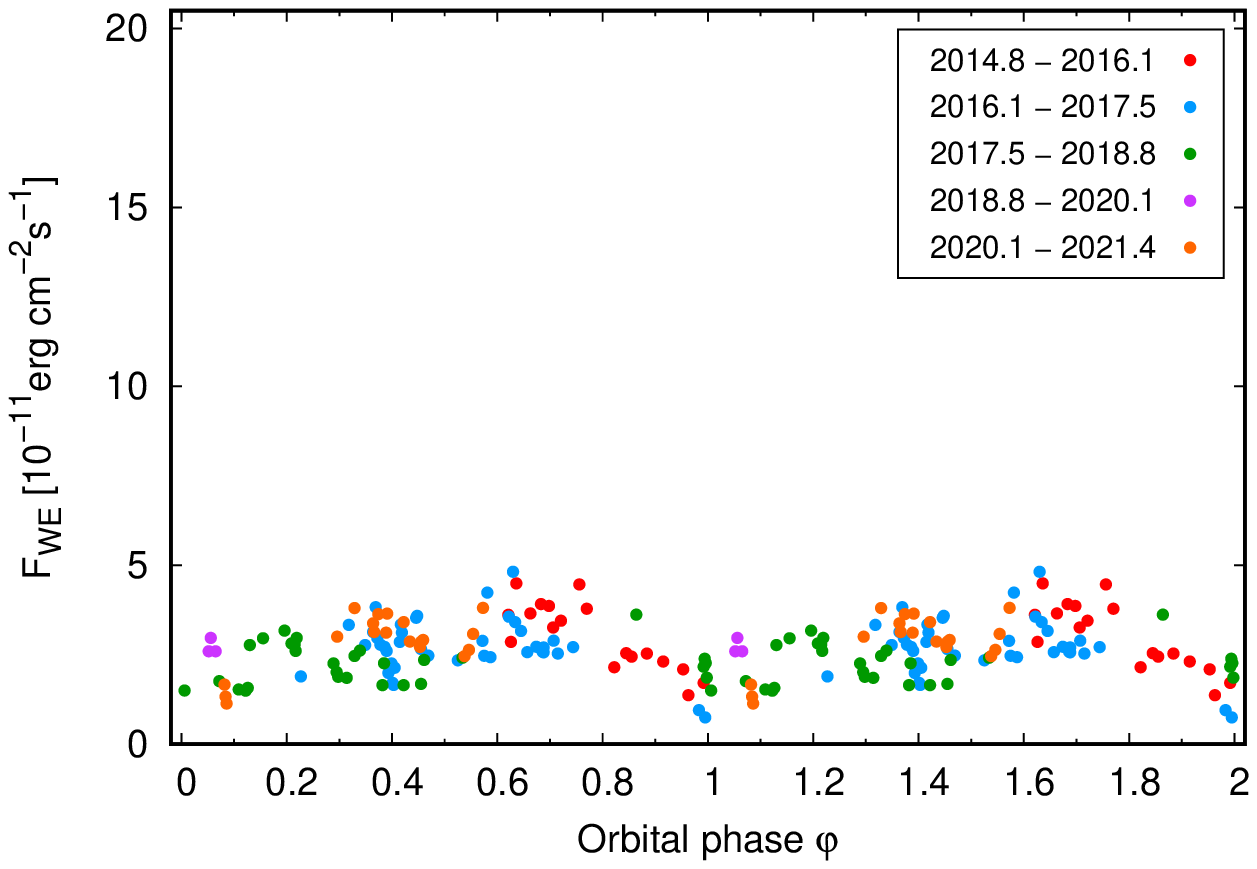}
          \includegraphics[angle=0]{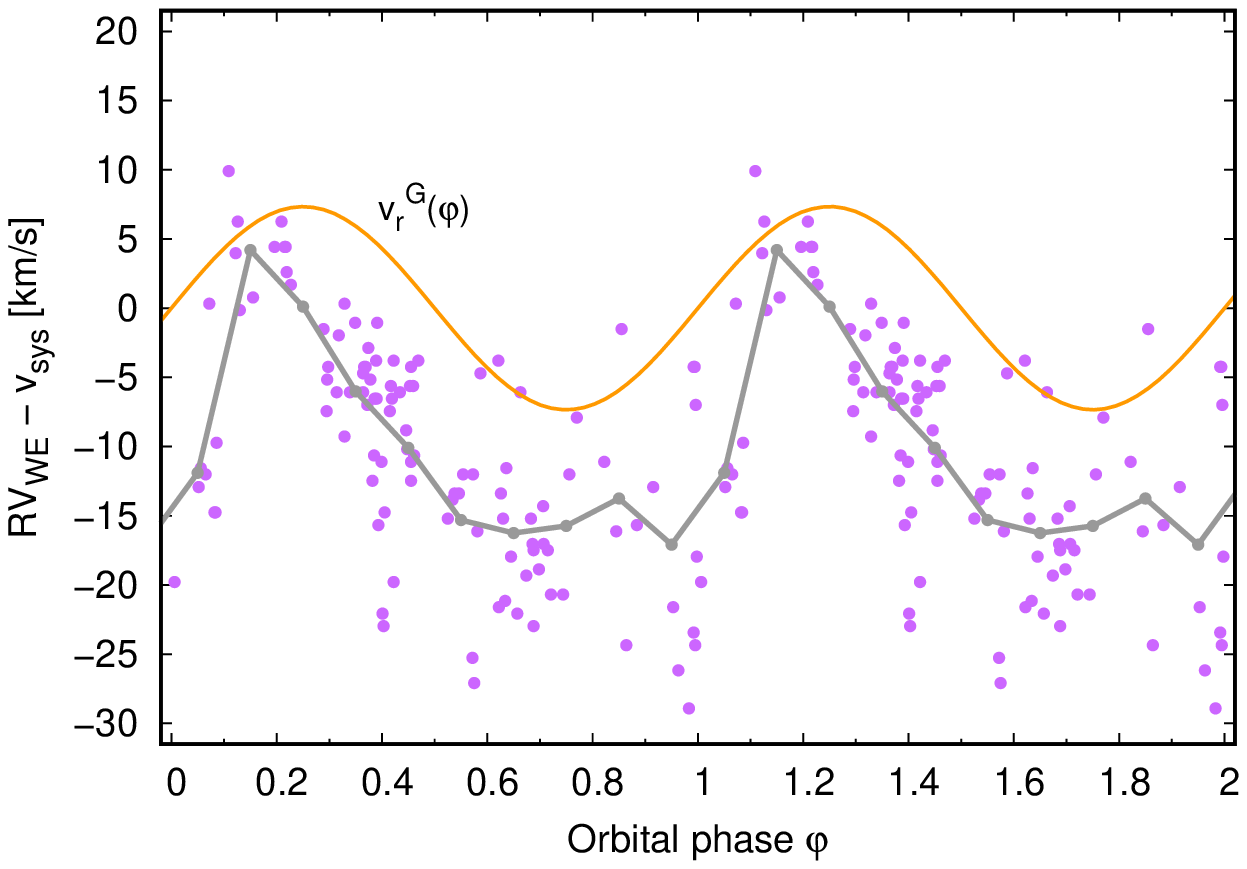}}
\end{center}
\caption{
Phase diagrams for fluxes (left) and RVs (right) of individual 
components of the \ha\ line profile; the CE (top), A (middle) 
and broad WE (bottom). Different orbital cycles 
are denoted by different colours (see the keys).
The orange line represents the RV-curve 
of the RG, $v_r^G(\varphi)$. The arrow in the top-left panel 
marks the brightening, possibly caused by the refraction effect 
(Sect.~\ref{ss:asymm}). 
The right-middle panel shows the shift of the A component 
(dashed line) relative to $v_r^G(\varphi),$ with the grey belt 
illustrating average uncertainty in measuring RVs 
(Sect.~\ref{sss:rvabs}). 
The grey line in the bottom-right panel represents binned RVs 
of the broad WE component with the bin width 0.1 of the orbital 
phase. Data are in Table~\ref{tab:par}.
        }
\label{fig:flx_rvs}
\end{figure*}

\begin{figure}[!]
\centering
\begin{center}
\resizebox{\hsize}{!}{\includegraphics[angle=0]
                     {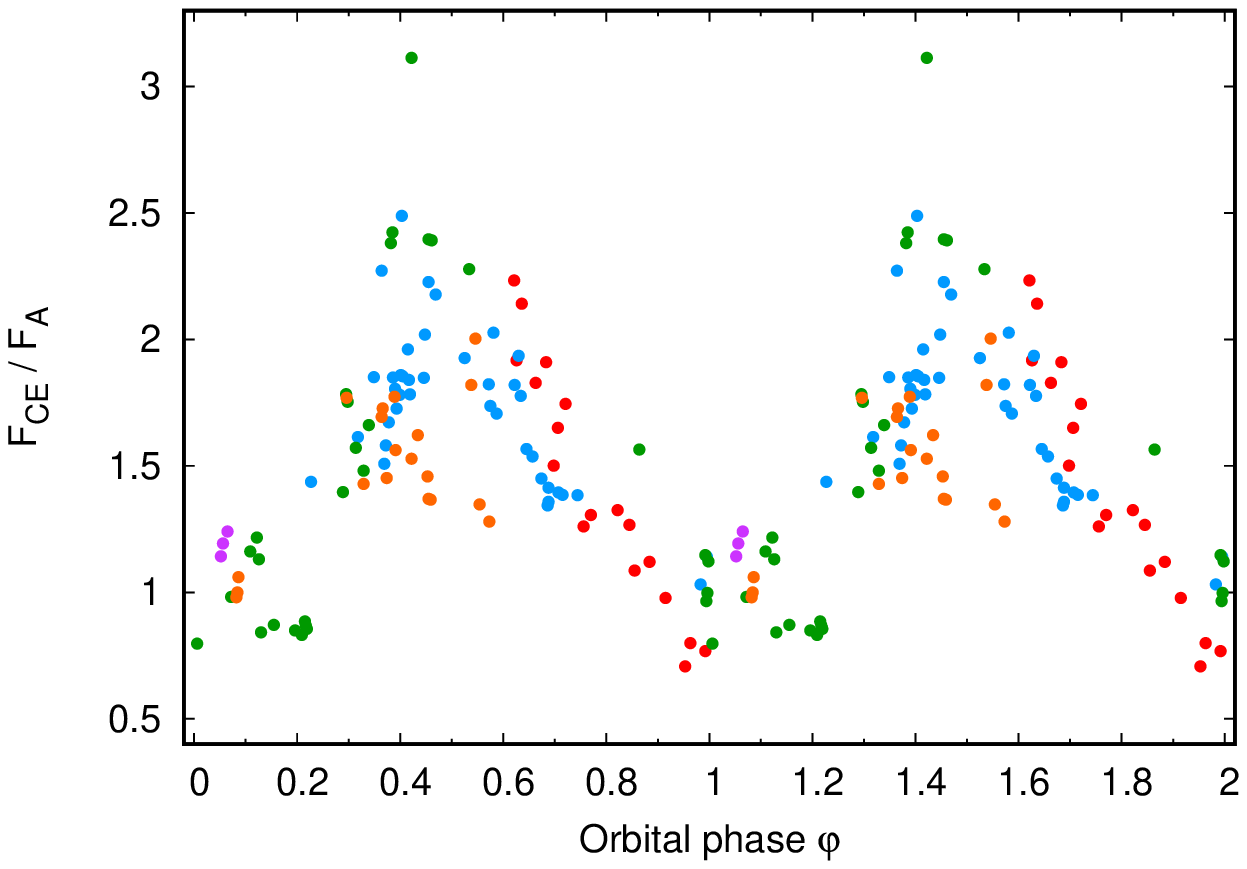}}
\end{center}
\caption{
As in Fig.~\ref{fig:flx_rvs}, but for flux ratio of the CE 
and A components. 
         }
\label{fig:fluxCEoverA}
\end{figure}

Measured fluxes of the \ha\ line components are plotted in 
Fig.~\ref{fig:flx_rvs} and summarised in Table~\ref{tab:par}. 
Fluxes of the CE component are strongest between 
$\varphi\sim 0.4$ and $\sim$0.7, followed by a broad 
asymmetric minimum up to $\varphi\sim 1.4$, where a steep 
increase towards the maximum is measured. This reflects an asymmetry 
of the wind density distribution with respect to the binary 
axis (see Sect.~\ref{ss:asymm}). 
Both CE and A components show a U-shaped profile between 
$\varphi = 0.4$ and $0.6$, suggesting an attenuation effect 
around the superior conjunction of the RG.  
Its origin probably inheres in a different opacity of the nebula 
at different orbital phases with a maximum around $\varphi = 0.5$, 
where the densest parts of the nebula located between the two 
stars are expected 
(see \cite{sk08} and Sect.~\ref{sss:focus1} of this paper).

A sign of the U-shaped feature around $\varphi = 0.5$ is also 
seen in the orbital evolution of the broad WE fluxes, while 
their minimum is at $\varphi \sim 0,$ implying the eclipsing 
nature of their source. 
Therefore, a substantial fraction of the broad WE source has 
to be located at/around the orbital plane, between the WD and 
the RG (see Sect.~\ref{sss:rvem}). 
Further interesting feature is the secondary maximum of the CE 
fluxes at $\varphi\approx 0.1$. We speculate that the nebular 
light generated at/above the top of the asymmetric H$^0$ conus 
can be enhanced by the refraction effect within the dense 
H$^0$ zone (see Sect.~\ref{ss:asymm}). 

Finally, the fact that we measure a residual emission in the \ha\ 
line around $\varphi = 0$ is a natural consequence of the ionisation 
structure of the binary, where a major part of the RG wind is 
ionised by the hard radiation from the WD (see Fig.~\ref{fig:h0h+}). 
%
%
%
\subsection{Radial velocities of the \ha\ line components}
\label{radvel}
Figure~\ref{fig:flx_rvs} shows the RVs of individual 
\ha\ line components 
with subtracted $v_{\rm sys}$ as a function of the orbital phase 
$\varphi$. For comparison, we plotted the RV curve of the RG, 
$v_r^G(\varphi)$, according to \cite{kg16}, as the only well-known 
motion of a source of radiation in the system. 

\subsubsection{The absorption component}
\label{sss:rvabs}
Radial velocities of the absorption component basically follow 
the orbital motion of the RG shifted by -7.5$\pm 2.5$\kms\ 
(Fig.~\ref{fig:flx_rvs}, middle right). 
Such a behaviour suggests that this component is created 
in the vicinity of the RG, within its wind. To estimate 
the height above the RG photosphere (i.e. the beginning of the 
wind), up to which the observer can see in the absorption core, 
we considered the wind velocity profile of \cite{sh+16}: 
\begin{equation}
   v(r) = \displaystyle\frac{v_\infty}{1 + \xi r^{1-K}},
\label{eq:wvp}
\end{equation}
where $r$ is the distance from the RG centre in units of its 
radius $R_{\rm G}$. The parameter of the wind model $\xi$ 
determines the distance from the RG surface at which the wind 
starts to accelerate significantly, while the parameter $K$ 
defines the steepness of this acceleration 
\citep[see Figs.~1 and 2 of][]{sh15}.
Then, for $R_{\rm G} = 75$\ro\ \citep[][]{vo+92}, and $\xi$ and $K$ 
for the orbital inclination $i = 80^\circ \pm 10^\circ$ 
\citep[see Table~3 of][]{sh+16}, a typical 
terminal velocity of the wind, $v_\infty = 30$\kms, the measured 
RV of the A component relative to the RG, $v_A = 7.5\pm 2.5$\kms, 
and the radial flow, the outer observer can see up to the height 
\begin{equation}
   R_A = 1^{+0.4}_{-0.3}\, R_{\rm G} 
\label{eq:ra}
\end{equation}
above the RG photosphere. 
For higher $v_A =10\pm 2.5$\kms, as measured around 
$\varphi = 0.2$ and $0.6-0.7$, the $R_A$ value is consistent 
with that in Eq.~\ref{eq:ra} within their uncertainties. 
%
%
\begin{figure}[]
\centering
\begin{center}
\resizebox{\hsize}{!}{\includegraphics[angle=0]{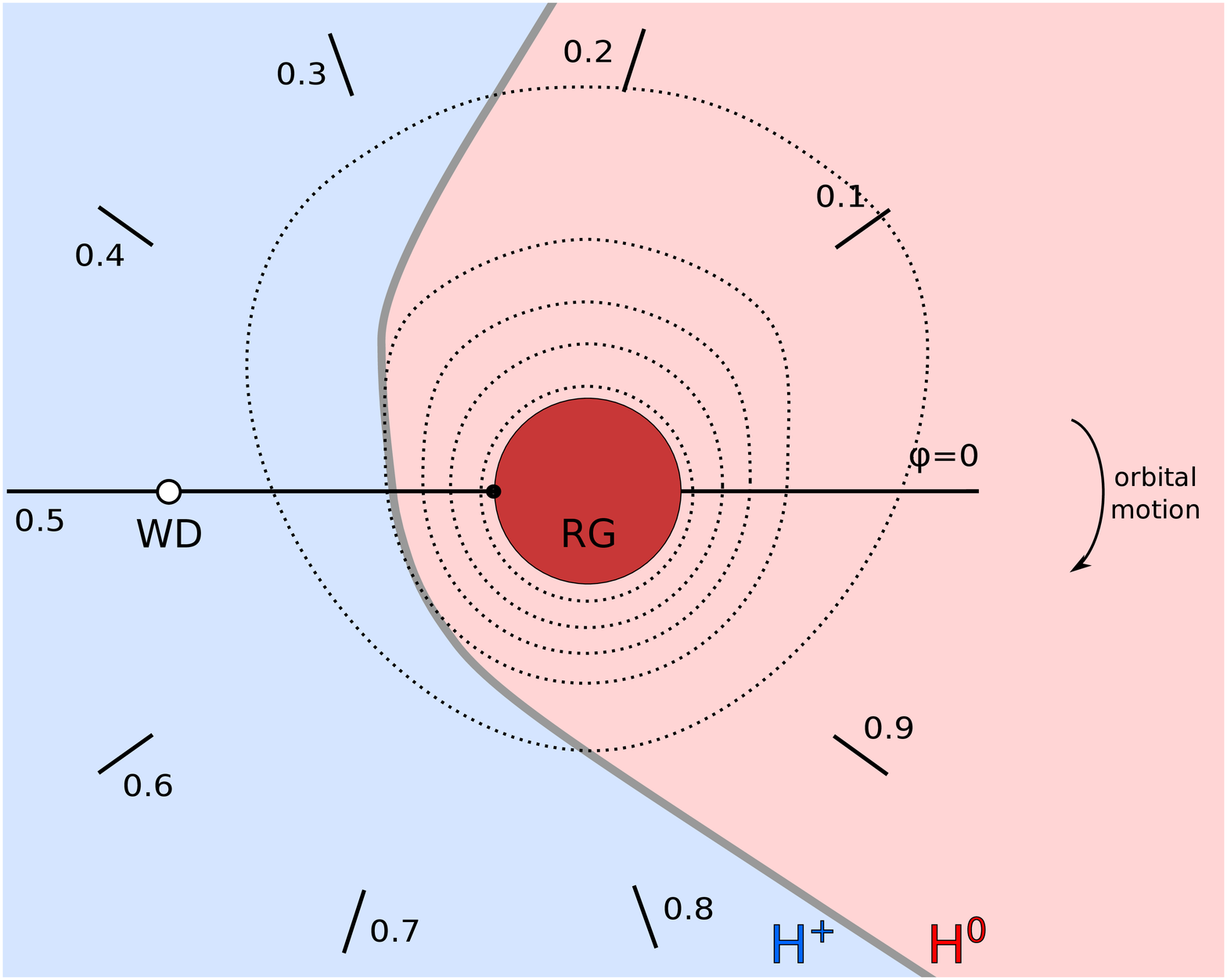}}
\end{center}
\caption{
Sketch of the basic ionisation structure for the symbiotic 
binary \object{EG And} in the orbital plane. 
The shape of the H$^{0}$/H$^{+}$ boundary and isodensities 
(dotted lines) are shown here for another quiet eclipsing 
symbiotic star \object{SY Mus} \citep[see][]{ss17,2017PASP..129f7001S} 
because of the lack of measured H$^0$ column densities at 
ingress phases for \object{EG And}.
The lines of sight point to the centre
of mass of the binary star (black point).
Isodensities demonstrate the asymmetric density distribution
of the giant's wind in the orbital plane.
In particular, a higher density gradient in front of the RG
in the direction of its orbital motion than behind
it (Sect.~\ref{ss:asymm}). 
         }
\label{fig:h0h+}
\end{figure}

\subsubsection{Emission components}
\label{sss:rvem}
The most notable feature of the CE RVs is their double-wave 
variation along the orbit (Fig.~\ref{fig:flx_rvs}, top right). 
From $\varphi\sim 0.4$ to $\sim 0.7$, they are scattered around 
the $v_r^G(\varphi)$ curve, while between $\varphi\sim 0.8$ and 
$\sim 1.3$ they are at its opposition. 

Since around $\varphi = 0.5$ we can see the densest parts of 
the nebula located at/above the ionisation boundary between 
the binary components, RVs of the CE component will obey the radial motion of the RG. 
At positions of $\varphi\sim$\,0.8--1.2, these parts of the 
nebula are occulted by the RG and its neutral wind zone. 
As a result, we can see a more distant nebular region better 
from/around the accreting WD. That explains why the RVs of 
the CE component are placed at the anti-phase of the 
$v_r^G$ curve at these phases (Fig.~\ref{fig:flx_rvs}). 
Around $\varphi = 0.6-07$, the CE RVs are systematically shifted 
by $\approx -5$\kms\ relative to the $v_r^G$ curve. This fact can 
reflect an enhanced flow from the RG as suggested by recent 
hydrodynamical modelling 
\citep[see Fig.~9 of][]{em+20}. 
%

Broad WE components are low with a flat maximum and a 
terminal velocity of $\sim$200\kms\ (Fig.~\ref{fitsHa}). 
Therefore, determination of their position is rather uncertain. 
Nevertheless, there is a redward-shifted maximum with 
RVs$\gtrsim -5$\kms\ at $\varphi = 0.1-0.3$, while within 
the rest part of the orbit we measure a broad minimum with 
RVs scattered from $\sim$-5 to $\sim$-25\kms. A relation to 
the $v_r^G$ curve shifted by $\approx$-10\kms\ can be recognised 
(see Fig.~\ref{fig:flx_rvs}, bottom right). This fact and the 
attenuation of this component around $\varphi = 0$ suggest that 
it is produced by Raman scattering of Ly-$\beta$ photons 
from/around the accreting WD on neutral atoms of hydrogen 
in the densest parts of the H$^0$ zone \citep[see][]{le00}. 
In this case, the H$^0$ atoms that move against the Ly-$\beta$ 
photons, predominantly along the binary axis, produce the blueshifted Raman photons, while scattering by receding H$^0$ atoms 
in the outer wind produce the redshifted 
Raman emission.\footnote{In this way, \cite{1999A&A...348..950S} 
explained the blue adjacent component in the Raman scattered 
O\,{\small VI}\,$\lambda$6825 profile in some SySts.} 
The blue Raman wing dominates the profile, because of the much 
higher density of the wind along the binary axis than in outer 
parts of the wind. 
Raman contributions from the neutral wind between the RG and 
the WD are therefore occulted around $\varphi = 0$. 
The systematic blueward shift of the broad WE component is 
partly compensated around $\varphi = 0.1-0.3$, where the main 
scattering region is drifted from the observer by the orbital 
motion of the RG (see Fig.~\ref{fig:h0h+}). 
\begin{figure}
\centering
\begin{center}
\resizebox{\hsize}{!}{\includegraphics[angle=-90]
                     {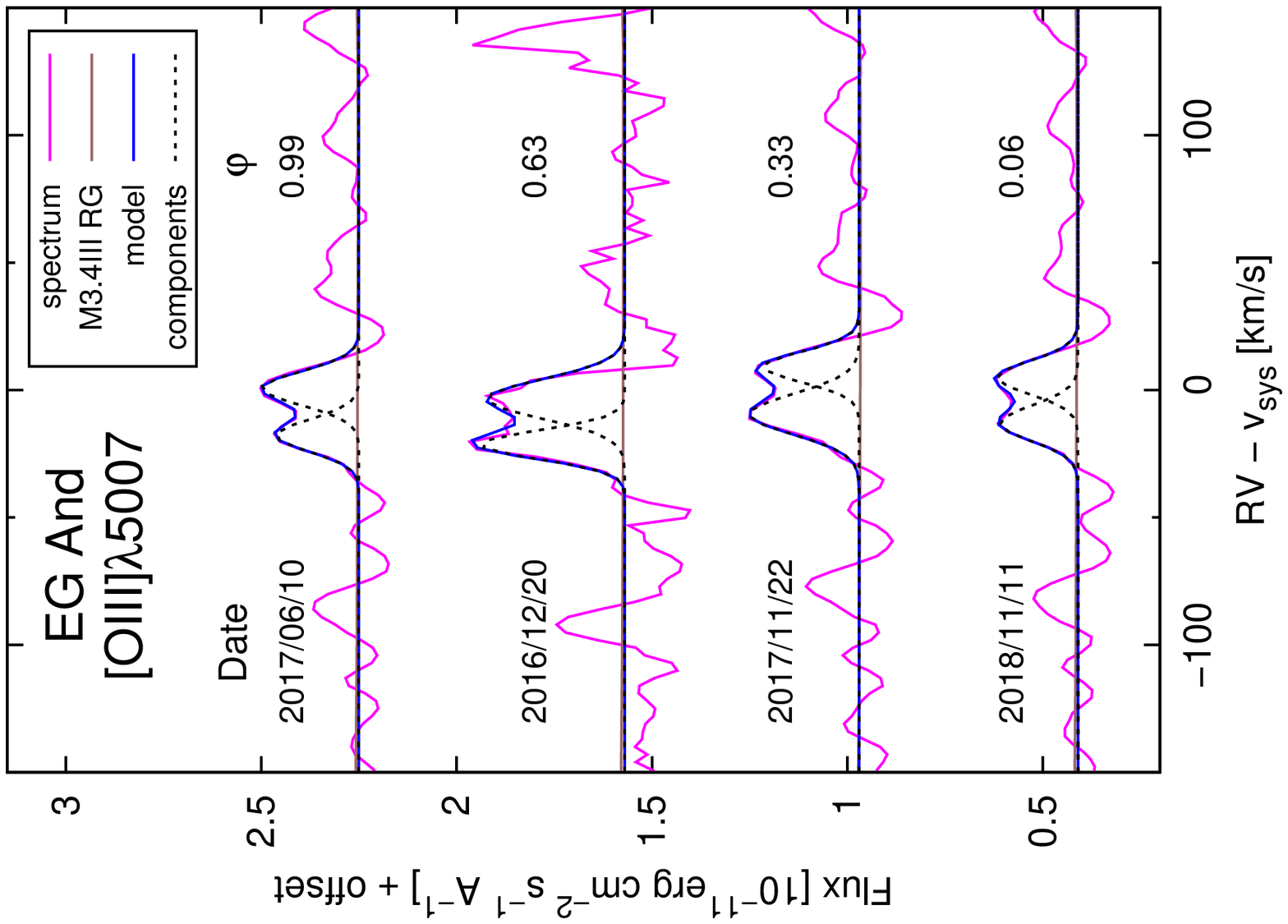}}
\end{center}
\caption{
As in Fig.~\ref{fitsHa}, but for [\ion{O}{iii}]\,$\lambda$5007 
line profile from our high-resolution spectra with the offset 
of $2\times\varphi$ for better visualisation. 
         }
\label{fig:oiii}
\end{figure}

\begin{table}[t!]
   \caption{
Radial velocities $RV_{\rm b}$ and $RV_{\rm r}$ of the blue
and red components of the \oIII\,$\lambda$5007~line profile 
measured on our high-resolution spectra.
           }
\label{tab:oiii}
\centering
\begin{tabular}{cccc}
\hline
\hline
\noalign{\smallskip}
Date                & 
Phase               &
$RV_{\rm b}$       & 
$RV_{\rm r}$       \\
(yyyy/mm/dd.ddd)                     &
                                     &
\multicolumn{2}{c}{[\kms]}           \\
\noalign{\smallskip}
\hline
\noalign{\smallskip}
 2016/12/20.939 & 0.630 & -116.34 & -98.38  \\
 2017/06/10.008 & 0.985 & -112.75 & -94.19  \\
 2017/06/20.012 & 0.006 & -110.35 & -92.39  \\
 2017/06/22.023 & 0.010 & -110.35 & -92.39  \\
 2017/07/07.017 & 0.041 & -109.15 & -91.79  \\
 2017/07/30.049 & 0.089 & -106.76 & -89.40  \\
 2017/11/15.921 & 0.314 & -104.36 & -85.80  \\
 2017/11/22.887 & 0.329 & -104.36 & -85.80  \\
 2018/09/09.058 & 0.930 & -114.54 & -95.98  \\
 2018/09/20.972 & 0.955 & -112.75 & -94.19  \\
 2018/11/11.850 & 0.062 & -107.96 & -89.99  \\
 2020/02/20.710 & 0.028 & -107.36 & -89.40  \\
 2020/03/02.735 & 0.050 & -107.96 & -89.99  \\
 2020/03/05.742 & 0.057 & -106.76 & -88.20  \\
 2020/03/09.738 & 0.065 & -105.56 & -87.00  \\
 2020/03/17.748 & 0.082 & -103.77 & -85.80  \\
 2020/03/18.743 & 0.084 & -104.36 & -85.80  \\
 2020/03/19.746 & 0.086 & -104.36 & -85.80  \\
 2020/06/29.042 & 0.296 & -103.77 & -85.20  \\
 2020/07/15.038 & 0.329 & -103.77 & -85.20  \\
 2020/07/31.961 & 0.364 & -104.36 & -85.80  \\ 
 2020/08/02.021 & 0.366 & -105.56 & -86.40  \\ 
 2020/08/05.985 & 0.374 & -106.16 & -87.00  \\ 
 2020/08/13.047 & 0.389 & -105.56 & -87.00  \\ 
 2020/08/14.006 & 0.391 & -105.56 & -87.00  \\ 
 2020/08/28.873 & 0.422 & -104.96 & -87.00  \\ 
 2020/09/03.986 & 0.434 & -106.16 & -87.60  \\ 
 2020/09/13.008 & 0.453 & -107.96 & -89.99  \\ 
 2020/09/13.915 & 0.455 & -109.15 & -89.99  \\ 
 2020/09/15.912 & 0.459 & -109.15 & -89.99  \\ 
 2020/10/27.911 & 0.546 & -113.35 & -93.59  \\
 2020/10/31.882 & 0.554 & -113.35 & -93.59  \\
 2020/11/09.805 & 0.573 & -113.35 & -93.59  \\
\noalign{\smallskip}                         
\hline                                      
\end{tabular}
\end{table}

\subsection{Orbital variability of the 
           [\ion{O}{iii}]\,$\lambda$5007 line}
\label{ss:oiii}
Our high-resolution spectra from the Skalnat\'e Pleso observatory 
($R\sim 38\,000$) revealed for the first time that the 
[\ion{O}{iii}]\,$\lambda 5007$ line consists of two components. Both 
are rather narrow; no broad wing emission is detectable 
(see Fig.~\ref{fig:oiii}). Therefore, to obtain information on 
the variability of the [\ion{O}{iii}] line, we fitted its profile 
by two Gaussian curves. The most striking variability is that 
in RVs (Table~\ref{tab:oiii}). Both the blue and the red 
component follow the RV curve 
of the RG shifted by $\sim -15$ and $\sim +3.0$\kms, respectively 
(Fig.~\ref{fig:focus}). 
Since the coverage of the orbital cycle with our high-resolution 
spectra is sparse, we also measured the average RV of the 
unresolved [\ion{O}{iii}]\,$\lambda$5007 line on the 
medium-resolution spectra. In this way, we verified the 
correlation between the RV curve of the RG and the RVs of the 
[\ion{O}{iii}]\,$\lambda$5007 line (Fig.~\ref{fig:focus}, top). 
A scatter in the [\ion{O}{iii}]-line RVs 
around $\varphi = 0.6-0.7$ is also present in RVs of the A and 
CE components in the \ha\ profile (Fig.~\ref{fig:flx_rvs}). 
This suggests the presence of a higher velocity and variable 
wind from the RG at these positions. 

%
%
\section{Discussion}
\label{s:dis}
\subsection{The asymmetry of the neutral wind zone}
\label{ss:asymm}
The presence of the neutral hydrogen region in SySts 
is directly proved by the Rayleigh scattering of the hot 
star continuum around the Ly-$\alpha$ line on neutral atoms of 
hydrogen \citep[e.g.][]{is+89}. 
The effect allows us to measure the hydrogen column density, 
$n_{\rm H}$, as a function of the orbital phase. In this way, 
\cite{du+99} presented evidence for an asymmetric 
density distribution around the RG in the eclipsing symbiotic 
binary \object{SY Mus} by comparing the egress and ingress 
$n_{\rm H}$ values. 
Modelling $n_{\rm H}(\varphi)$ values measured for 
\object{SY Mus}, \cite{sh+16} derived the wind velocity profiles that 
require lower number density of the wind in front of the RG 
motion than behind it at distances $\gtrsim 1\,R_{\rm G}$ above 
the RG surface
(see Fig.~1 of \cite{2017PASP..129f7001S} and Fig.~\ref{fig:h0h+} here).
As a result, the H$^{0}$/H$^{+}$ ionisation boundary is shaped 
asymmetrically with respect to the binary axis at the orbital 
plane \citep[see Fig.~3 of][]{ss17}. 
Our observations support the asymmetric shaping of the 
H$^{0}$ zone in \object{EG And} as follows.

Firstly, asymmetric distribution of the CE flux showing the highest 
values between $\varphi\sim$0.4 and $\sim$0.7, where its RVs 
follow the $v_r^G(\varphi)$ curve (see Sect.~\ref{sss:rvem}), 
and the gradual decrease until $\varphi \sim 0.9$, after which 
the densest parts of the nebula are not visible, are consistent 
with the asymmetric shaping of the H$^{0}$/H$^{+}$ boundary. Secondly, supposing that the shape of the ionisation boundary and 
isodensities are similar in quiet S-type symbiotic stars 
(Fig.~\ref{fig:h0h+}), the sharp increase of the CE flux 
between $\varphi = 0.3$ and 0.4 by a factor of 2--3 
(Fig.~\ref{fig:flx_rvs}) can be explained by uncovering 
the dense parts of the ionised wind located 
in the orbital plane, just above the ionisation boundary, 
approximately in the direction of the RG orbital motion. Thirdly, higher fluxes of the broad WE component measured between 
$\varphi\sim 0.6$ and $\sim$0.9 with a minimum around $\varphi = 0$ 
(Fig.~\ref{fig:flx_rvs}), and their origin within the densest 
parts of the neutral wind by Raman scattering 
(see Sect.~\ref{sss:rvem}), are also consistent with the asymmetric 
H$^0$ density distribution around the RG.

Finally, a fraction of the CE flux generated in 
the H$^{+}$ zone enters the H$^{0}$ zone, where it can be 
refracted and thus creates a local light maximum 
around $\varphi = 0.1$ (Fig.~\ref{fig:flx_rvs}). 
Its position after the time of the RG inferior conjunction 
independently confirms the asymmetric wind density distribution 
around the RG, primarily derived from Rayleigh scattering 
(see above). 
A relatively narrow peak of the brightening can be a result of 
focusing a monochromatic light 
of the \ha\ line, while its rather broad base 
consists of contributions from a dense part of 
the nebula spread just above the H$^{0}$/H$^{+}$ boundary, 
around the binary axis. 
\begin{figure}[!t]
\centering
\begin{center}
\resizebox{\hsize}{!}{\includegraphics[angle=0]
                        {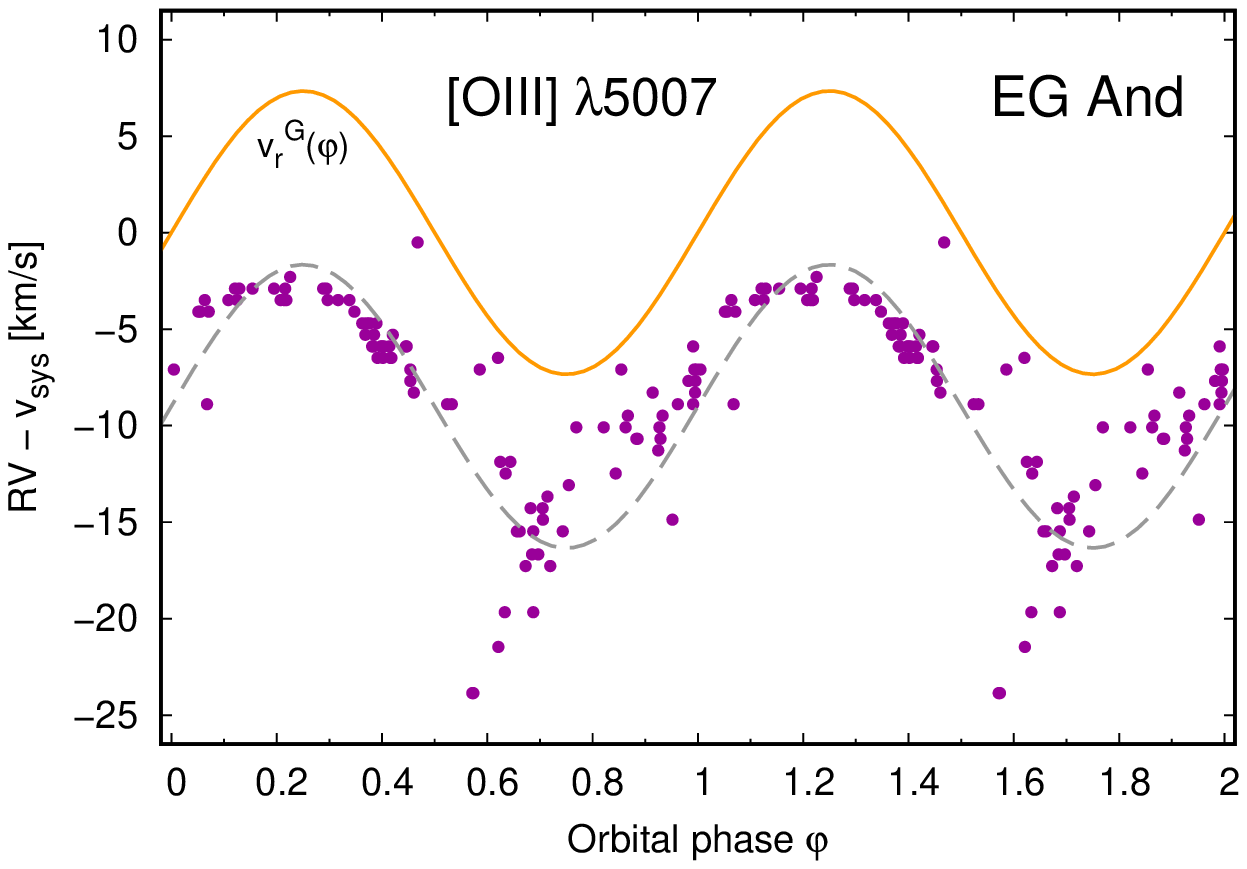}}
\resizebox{\hsize}{!}{\includegraphics[angle=0]
                       {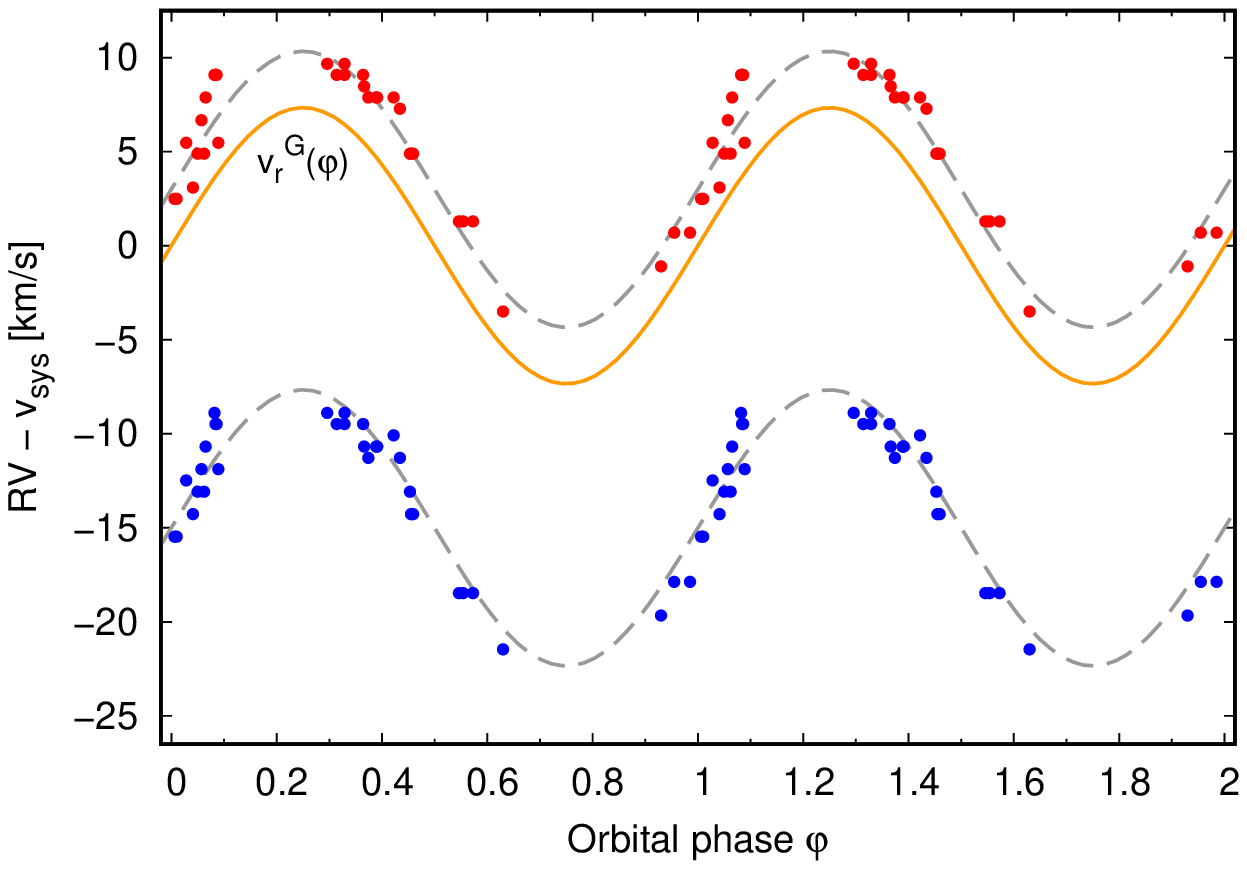}}
\resizebox{\hsize}{!}{\includegraphics[angle=0]
                                 {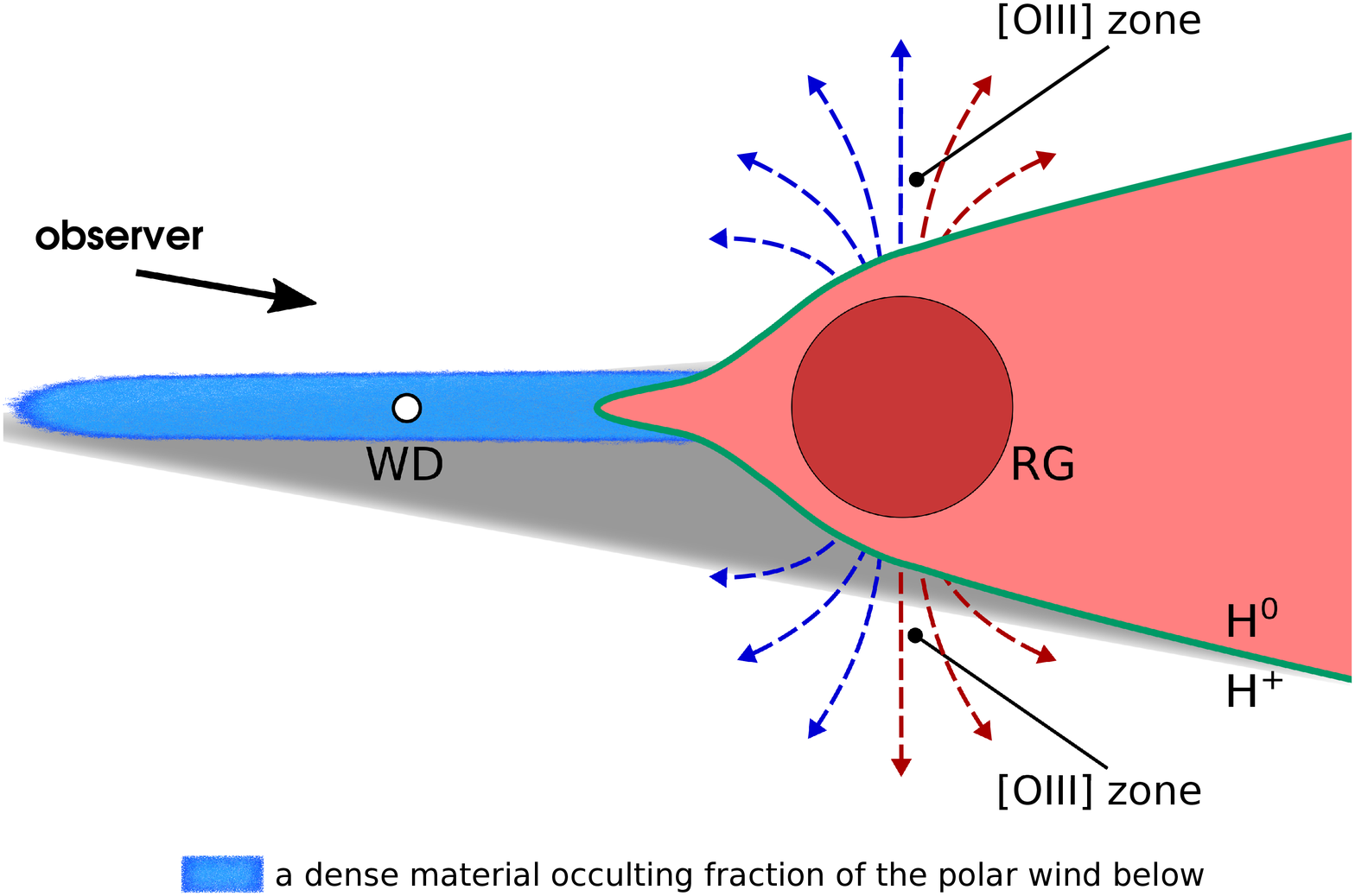}}
\end{center}
\caption{
{\it Top}: RVs of [\ion{O}{iii}]\,$\lambda$5007 line from 
medium-resolution spectra. They are shifted by -9\kms\ (grey 
dashed line) from the RG motion (orange line). 
{\it Middle}: RVs of blue and red components of the profile 
from high-resolution spectra. Both components follow the RG 
orbital motion, being shifted by -15 and +3.0\kms\ (grey dashed 
lines). 
{\it Bottom}: 
Sketch of [\ion{O}{iii}] line zones in the vicinity of the RG 
at/around its polar regions as seen on the cut perpendicular to 
the orbital plane containing the binary axis 
(see Sect.~\ref{sss:focus2}). 
The H$^0$/H$^+$ boundary is estimated according to the wind 
compression disk model \citep[see Fig.~2 of][]{2015A&A...573A...8S}. 
         }
\label{fig:focus}
\end{figure}
%
\subsection{The wind focusing towards the orbital plane}
\label{ss:focus}
Theoretical calculations showed that the presence of a companion 
reduces the effective gravity of the mass-losing star in the binary, 
which leads to the enhancement of the wind mass loss in the orbital plane 
\citep[][]{1988MNRAS.231..823T,2001A&A...367..513F,mp07,vb+09,
vb+17,2012A&A...542A..42H,bo+16}. 
For S-type SySts, a slow rotation of their normal RGs can also 
contribute to high compression of the wind towards the orbital 
plane \citep[][]{2015A&A...573A...8S}. 

The first observational indication of the wind focusing from 
the RG in S-type SySt was suggested for \object{SS Leporis} by 
interferometric measurements. 
It was found that the RG is well below its Roche lobe. 
Consequently, an enhanced wind mass loss 
in the orbital plane is required, which fuels
the abnormal luminosity of the hot component 
\citep[][]{2011A&A...536A..55B}. 
Afterwards, modelling the $n_{\rm H}(\varphi)$ values for 
quiescent S-type systems \object{EG And} and \object{SY Mus}, \cite{sh+16} 
derived a factor of $\ga$10 higher mass-loss rates in 
the near-orbital-plane region than the total rate derived 
from nebular emission. Here, our observations indicate the wind 
focusing in \object{EG And} from the RVs of the nebular 
[\ion{O}{iii}]\,$\lambda$5007 line, independently 
of the same finding from the column density models 
of \cite{sh+16}. 
\subsubsection{Focusing from the [\ion{O}{iii}]\,$\lambda$5007 line}
\label{sss:focus2}
Figure~\ref{fig:oiii} shows example of the [\ion{O}{iii}]\,$\lambda$5007 
line profiles seen on our high-resolution spectra. 
The fact that the profile consists of two individual components 
implies the presence of two detached emitting regions. 
No shift in the orbital phase and almost no difference in the 
amplitude of their RVs with respect to the $v_r^G(\varphi)$ 
curve imply that both regions are located in the vicinity of 
the RG (Fig.~\ref{fig:focus}). 
The [\ion{O}{iii}]\,$\lambda$5007 line arises in the O$^{++}$ ion by 
transitions from the metastable $^{1}D_2$ to the ground 
$^{3}P_2$ level. This line can arise only within the H$^{+}$ 
zone, where photons capable of creating the O$^{++}$ ions 
($h\nu > 35$\,eV) can be present. 
These forbidden transitions can create a strong emission in 
extended low density nebulae with the electron concentration 
$n_{\rm e}$ of the order of $10^3 - 10^4$\cmt. 
At higher densities, a deactivation of nebular transitions 
mainly by free electrons weakens this line 
\citep[e.g.][]{1997pdpn.book.....G}. Using the corresponding 
deactivation factor, \cite{2001A&A...367..199S} found that 
faint [\ion{O}{iii}]\ nebular lines in the \object{AX Per} spectrum are created 
in the region with $n_{\rm e} \approx 3\times10^7$\cmt, and they 
derived the upper limit of 
$n_{\rm e}([\ion{O}{iii}])$ as $7\times 10^7$\cmt. 

The limiting density for emergence of the nebular [\ion{O}{iii}]\ 
lines constrains the density distribution of the wind in the 
vicinity of the RG. The spherically symmetric wind driven at 
rates of $\approx 10^{-7}$\myr, as derived from nebular emission 
by \cite{1993ApJ...410..260S}\footnote{The radio emission 
was measured around $\varphi =0$ and the distance $d=0.31$\,kpc was 
used to derive $\dot{M}_{\rm G}\approx 10^{-8}$\myr. Estimating 
the radio emission for $\varphi\sim 0.5$ from the total emission 
measure at this phase \citep[see Table~3 of][]{sk05} yields 
$\dot{M}_{\rm G}\approx 10^{-7}$\myr\ for $d=0.59$\,kpc.} and 
\cite{sk05}, corresponds to particle densities $> 10^8$\cmt\ 
up to the radial distance from the RG centre $r \sim 5\,R_{\rm G}$ 
(for $v(r)$ given by Eq.~(\ref{eq:wvp}) and $v_\infty = 30$\kms), 
which is comparable to the separation between the binary 
components \citep[][]{kg16}. 
However, the location of the nebular [\ion{O}{iii}] line region 
in the vicinity of the RG requires the mass-loss rate of 
$\lesssim 10^{-8}$\myr\ to achieve concentrations of a few times 
$10^7$\cmt\ at $r \approx 2-3\,R_{\rm G}$. 
Therefore, the wind from the RG in \object{EG And} has to be substantially 
compressed towards the orbital plane to satisfy: 
(i) the rates of a few times $10^{-6}$\myr\ measured in the 
near-orbital-plane region from Rayleigh scattering \citep[][]{sh+16}; 
(ii) the average rates of $\approx 10^{-7}$\myr\ derived from 
nebular emission of the ionised wind; and 
(iii) the rates of $\lesssim 10^{-8}$\myr\ from the polar regions 
required by the location of the [\ion{O}{iii}]\,$\lambda$5007 line zones 
in the vicinity of the RG. 

Figure~\ref{fig:focus} shows a sketch for the [\ion{O}{iii}] line zones 
placed in the vicinity of the RG as allowed by the wind compression 
towards the orbital plane due to its rotation  
\citep[see Fig.~2 of][]{2015A&A...573A...8S}.
 Due to the wind compression, the H$^{+}$ region is 
close to the RG photosphere around its poles and at the 
neighbouring parts with the diluted wind facing 
the hot component. 
Then, viewing the binary under the inclination angle of 
$\sim 80^{\circ}$, trajectories of the polar wind 
from the above giant's hemisphere are directed more towards the observer, 
while those from the opposite hemisphere are directed 
away from the observer.
Therefore, the wind from the polar emission zone above 
the orbital plane contributes mainly to the blueshifted 
component of the line, whereas the opposite polar emission 
zone is responsible for the red component.  
In addition, a fraction of the polar wind below the orbital
plane can be occulted by the dense orbital-plane material 
(Fig.~\ref{fig:focus}).
This is supported by the fact that the blue component of 
the [\ion{O}{iii}]\,$\lambda$5007 line is shifted more 
from the $v_r^G(\varphi)$ curve, while the red one is 
close to it.
%
%
%
\subsubsection{CE and A components from wind focusing}
\label{sss:focus1}
Orbitally related variation of the CE fluxes show a factor of 
$\sim$5 difference between the maximum and minimum.  
The maximum is more or less rectangular in profile, with a 
base of $\Delta\varphi\sim 0.4 - 0.65$, followed with a broad 
minimum lasting from $\varphi\sim 0.7$ to $\varphi\sim 1.4$ 
(see Fig.~\ref{fig:flx_rvs}). 
Such a profile suggests that a major part of the CE flux 
produced by the H$^{+}$ zone is occulted by the RG and/or 
attenuated by the neutral wind zone at these positions. 
As the H$^{0}$ zone is flattened towards the orbital plane 
due to the dilution of the polar wind, a significant part 
of the emitting material has to be more concentrated 
on the orbital plane. 
%

The maximum of the absorption component is 
observed around the superior conjunction of the RG, from 
$\varphi\sim 0.4$ to $\sim$0.65, instead of the expected 
opposite position around 
$\varphi = 0$ (see Fig.~\ref{fig:flx_rvs}). 
This anomaly, a correlation with the CE flux, can be 
explained by an attenuation of \ha\ photons within both 
the neutral and the ionised zone around $\varphi = 0.5$ as follows: 
(i) The wind enhanced at the orbital plane makes the H$^{0}$ 
zone more dense here, shifts its boundary more 
to the source 
of ionising photons, and thus increases its optical depth. 
As a result, a stronger photospheric absorption within the 
H$^{0}$ wind region is created with respect to the spherically 
symmetric wind. 
(ii)
Assuming that the nebula is partially optically thick in 
\ha, a fraction of \ha\ photons produced by the H$^+$ 
zone are lost from the line of sight. 
The highest emissivity of the nebula is expected around 
$\varphi = 0.5$, where its largest volume containing densest 
parts above the ionisation boundary around the binary 
axis is seen. 
 The nebula simultaneously has the largest optical 
depth around this position, because of its extension here 
to the observer and concentration on the orbital 
plane.\footnote{That is, a shorter 
recombination time in a denser 
nebula increases the fraction of the H$^0$ atoms in 
the H$^+$ zone, which makes it more opaque for diffuse 
\ha\ photons.} 
As a result, a reduction of \ha\ photons 
on the line of sight 
from the H$^+$ zone is superposed with that caused by 
scattering in the \ha\ line within the H$^0$ zone. 
Therefore, the fit to the profile indicates the largest total 
decrease of \ha\ photons around $\varphi = 0.5$, 
and we measure 
a correlation between the CE flux and the flux in the A 
component (see Fig.~\ref{fig:flx_rvs}). 
\section{Summary}
\label{s:sum}
In this paper, we analyse the orbital variability of the \ha\ 
and [\ion{O}{iii}]\,$\lambda$5007 line profiles observed in 
the spectrum of the quiescent eclipsing SySt \object{EG And}. 
The main goal is to obtain information on the basic structure 
of the wind from the RG in this wide interacting binary. 
Our analysis revealed two pivotal characteristics of the wind: 
\begin{enumerate}
\item
The neutral wind zone is distributed asymmetrically at 
the orbital plane with respect to the binary axis. 
\item
The wind from the RG is substantially compressed to 
the orbital plane from polar directions. 
\end{enumerate}
These conclusions are supported by the variation of fluxes 
and RVs of individual components of the \ha\ and 
[\ion{O}{iii}]\,$\lambda$5007 line profiles along the orbit 
(see Figs.~\ref{fig:flx_rvs} and \ref{fig:oiii}). The following 
points are also significant: 
\begin{itemize}
\item
Fluxes of the CE component are mostly produced by 
the densest region of the ionised wind located at/above the 
ionisation boundary facing the hot component 
(Sect.~\ref{sss:rvem}). 
Their asymmetric time evolution along the orbit 
(Sect.~\ref{ss:fha}, Fig.~\ref{fig:flx_rvs}) 
suggests the asymmetric shaping of the H$^{0}$/H$^{+}$ boundary 
at the orbital plane with respect to the binary axis. Also, the asymmetric wind density distribution around 
the RG is confirmed by the presence of the secondary maximum of 
the CE fluxes around $\varphi = 0.1$, which can result from 
refraction of the CE light in the neutral wind. Fluxes and RVs of the broad WE component, which originate from the densest parts of the neutral wind by Raman scattering, 
also reflect the asymmetric density distribution of H$^0$ atoms 
around the RG (Sect.~\ref{ss:asymm}). 
\item
A substantial compression of the RG wind to the orbital plane is 
constrained by the location of the nebular [\ion{O}{iii}] line 
zones in the vicinity of the RG ($r\approx 2-3\,R_{\rm G}$), 
at/around its poles. This requires the mass-loss rate of 
$\lesssim 10^{-8}$\myr\ from the polar regions 
(Sect.~\ref{sss:focus2},  Fig.~\ref{fig:focus}) to be two orders 
of magnitude lower than that measured in the near-orbital-plane 
region from Rayleigh scattering.  
\end{itemize}
Finally, we remark that the substantial focusing of the wind 
towards the orbital plane can mimic the ellipsoidal shape of 
the RG, and thus cause the double-wave profile 
of the optical light curves along the orbit even for systems 
containing RGs well within their Roche lobes -- a long-standing 
problem for the interpretation of this type of variability for \object{EG And} 
\citep[][]{1997MNRAS.291...54W,2001A&A...366..157S,kg16, 
2019CoSka..49...19S}. 
This problem is remarkable 
mainly for yellow SySts, whose G-K giants are deeply inside of 
their Roche lobes, but their optical light curves show 
a pronounced double-wave profile 
\citep[see Fig.~6.2 of][]{mu19}.

%

\begin{acknowledgements}
The authors are grateful to Matej Seker\' a\v s for obtaining 
a fraction of the spectroscopic observations used in this work 
and helpful discussions concerning the data reduction. 
Theodor Pribulla is acknowledged for the advice on 
verification of the wavelength calibration of the spectra.
Yolande Buchet, Zolt\' an Garai, Joan Guarro Flo, Tim Lester, 
Gerard Martineau, Theodor Pribulla, Peter Sivani\v c and Martin Va\v nko are 
thanked for obtaining 
1--2 spectra. Some spectra were acquired within the Astronomical 
Ring for Access to Spectroscopy (ARAS), an initiative promoting 
cooperation between professional and amateur astronomers in 
the field of spectroscopy, coordinated by Fran\c{c}ois Teyssier. 
This work was supported by the Slovak Research and Development
Agency under the contract No. APVV-15-0458 and by a grant of 
the Slovak Academy of Sciences, VEGA No. 2/0008/17. 
\end{acknowledgements}


\bibliographystyle{aa}

\bibliography{reflist.bib}

\end{document}